\documentclass[12pt]{article}
\usepackage[utf8]{inputenc}
\usepackage[left=1in, right=1in, top=1in]{geometry}
\usepackage[utf8]{inputenc}
\usepackage{appendix}
\usepackage{mathtools}
\usepackage{amsmath}
\usepackage{amssymb}
\usepackage{enumitem}
\usepackage{comment}
\usepackage{amsthm}
\usepackage{setspace}
\usepackage{graphicx}
\usepackage{color}
\usepackage{braket}
\usepackage{bbold}
\usepackage{url}
\usepackage{bibentry}
\usepackage{amsmath}
\usepackage{amsfonts}
\usepackage{clipboard}
\usepackage{cite}
\usepackage[colorlinks=true, linkcolor = blue, citecolor = blue, urlcolor = blue,linktocpage=true]{hyperref}
\usepackage{amssymb}
\usepackage{graphicx}
\usepackage{float}
\usepackage{caption}
\numberwithin{equation}{section}

\newcommand{\braout}{\bra{\mathrm{out}}}
\newcommand{\ketin}{\ket{\mathrm{in}}}
\newcommand{\odress}{e^{-R_{N}^{(\mathrm{out})}}}
\newcommand{\idress}{e^{R_{N}^{(\mathrm{in})}}}

\newcommand{\qout}{U_{E}^{\dagger\mathrm{out}}}
\newcommand{\qin}{U_{E}^{\mathrm{in}}}
\newcommand{\sa}{\mathrm{S}}
\newcommand{\sca}{\braout \mathrm{S} \ketin}
\newcommand{\qsa}{Q_{\mathrm{soft}}^{a}(\hat{{p_{1}}})}
\newcommand{\qsb}{Q_{\mathrm{soft}}^{b}(\hat{{p_{2}}})}

\newcommand{\qhb}{Q_{\mathrm{hard}}^{b}(\hat{{p_{2}}})}

\def\beq{\begin{equation}}
\def\eeq{\end{equation}}
\def\beqa{\begin{eqnarray}}
\def\eeqa{\end{eqnarray}}

\title{FK states in QCD}
\author{Anupam A H, Athira P V}
\date{April 2019}

\begin{document}

\baselineskip 24pt

\begin{center}

{\LARGE Generalised coherent States in QCD \\
	from asymptotic symmetries \par}

\end{center}

\vskip .5cm
\medskip

\vspace*{4.0ex}

\baselineskip=18pt

\centerline{\large \rm Anupam A H$^{a}$, Athira P V$^{b}$}

\vspace*{4.0ex}

\centerline{\it ~$^a$Institute of Mathematical Sciences, Taramani, Chennai 600 113, India}
\centerline{\it ~$^a$Homi Bhabha National Institute}

\vspace*{1.0ex}

\centerline{\it ~$^b$Chennai Mathematical Institute,  SIPCOT IT Park, Siruseri, Chennai, 603103 India} 

\vspace*{1.0ex}

\vspace*{1.0ex}
\centerline{\small \textit{E-mail :}  \href{mailto:anupam@imsc.res.in}{ \texttt{anupam@imsc.res.in}}, \href{mailto:athira@cmi.ac.in}{\texttt{athira@cmi.ac.in}}, }

\vspace*{5.0ex}

\centerline{\bf Abstract} \bigskip

We investigate the relationship between asymptotic symmetries of QCD and vacuum transitions induced during scattering.  Starting with the Fock vacuum in the far past, the infinity of conservation laws associated to (non-Abelian) asymptotic symmetries in QCD can be used to determine the vacuum in the far future. We show that the corresponding asymptotic Hilbert space in the future is generated by a class of dressed states in which each finite energy particle is dressed by a cloud of interacting soft gluons. We identify the precise structure of the dressing using conservation laws and show that the corresponding asymptotic states are equivalent to the generalised coherent states defined by Catani et al. 

\vfill \eject

\baselineskip=18pt

\tableofcontents

\section{Introduction}\label{intro}

Since the seminal work of Strominger in \cite{strom-ym1}, conservation laws associated to asymptotic symmetries have played an increasingly important role in uncovering the infra-red structure of gauge theories and gravity. The associated constraints on the S-matrix of these theories are intimately connected to soft theorems. For the so-called leading asymptotic symmetries such as large gauge transformations in Yang-Mills theory and super-translations in perturbative gravity,  these constraints relate the residue of the Weinberg soft factor to the insertion of soft modes in the scattering states of the theory.  This relationship between soft theorems and conservation laws has an interesting off-shoot \cite{strom-fadeev,sever,akhoury2,akhoury3,ashtekar} in that the insertion of soft modes results in a shifted vacuum. Due to an infinity of soft modes parametrized by their location on the celestial sphere, one has an infinite degeneracy of vacuua. It was shown in \cite{sever,strom-fadeev,akhoury2,akhoury3} that transitions between these vacua are constrained by  conservation laws and starting with a vacuum in the far past, conservation laws imply that scattering states in the far future are such that the finite energy particles are dressed by a cloud of low frequency modes. In QED and gravity it has been shown that these ``dressed states" are equivalent to the well known Faddeev-Kulish states. Thus starting with a number of technical assumptions primary among which is the infinity of conservation laws, one can obtain the dressed states of the theory in which S-matrix is infra-red finite.

We would like to ask precisely the same question in QCD. That is, starting with the infinite dimensional non-Abelian group of asymptotic symmetries, what does a given vacuum of perturbative QCD transit to in the far future. In a nutshell we show that a close scrutiny of existing works on the relationship between dressed states and symmetries in perturbative gravity helps us to distil a  minimal set of assumptions under which one obtains the asymptotic states in which finite energy partons are dressed by a cloud of (infinitely) correlated soft gluons. We argue that these states are equivalent to the well known dressed states in QCD \cite{greco,catani1}.

We emphasise that our analysis of determining the dressing only involves asymptotic symmetries and the so-called orthogonality condition. We do not use any multi-soft gluon theorem  (which is equivalent to Ward identities in un-dressed Fock states) to help us fix the dressing operator. This leads to a rather intricate analysis in which higher order (nested) Ward identities, orthogonality with respect to insertion of multiple gluons and global color conservation come together to give the definition of a dressed state.

The paper is organised as follows. In section \ref{gr} we recast the ideas in \cite{strom-fadeev,akhoury2,akhoury3,sever} to derive the Faddeev-Kulish states in gravity. In section \ref{qcd} we apply these ideas in the context of QCD so as to derive a set of dressed states in QCD. In section \ref{qcd4} we show that under a set of assumptions, one can recover the dressed states in QCD, which were obtained by Catani et al in \cite{catani1}, from Ward identities associated to asymptotic symmetries in QCD.

{\bf Note added} : While this manuscript was being prepared, a paper \cite{tristan} appeared which also investigates the relationship between dressed states in QCD and asymptotic symmetries. In that paper, starting from the generalised coherent states of Catani et al, the authors analyse the  quantum correction to the asymptotic charges as a result of such a dressing. Our work is rather complementary in the sense that we ask if conservation laws associated to asymptotic symmetries can be used to derive the nature of dressed states when aided by certain well motivated assumptions. In this sense we believe that the two papers complement each other.

\section{From Asymptotic Symmetries to Dressed States in Gravity}\label{gr}
In this section we revisit the works of \cite{akhoury1,akhoury2,akhoury3,sever,strom-fadeev} with a motive of understanding the dressed states in gravity and their relationship with asymptotic symmetries. We begin by reviewing the relationship between Ward identities associated to BMS supertranslations and leading single soft graviton theorem \cite{strom,alok:massive} in section \ref{gr1}. In section \ref{gr2} and \ref{gr3}, a brief review of the work \cite{akhoury3} in connecting the dressed states in gravity and BMS supertranslations is presented. Further in sections \ref{gr4} and \ref{gr5}, we show that under certain assumption one can derive the Faddeev-Kulish states in gravity from BMS supertranslations. In section \ref{gr6} an alternative asymptotic state to the Faddeev-Kulish state is presented.  

\subsection{Review of BMS Supertranslation Ward Identity and Leading Single Soft Graviton Theorem}\label{gr1}
We start with the definition of the supertranslation charge $Q[f]$ \cite{strom}. Supertranslation charges are associated to the class of vector fields that preserve the structure of an asymptotically flat metric. These charges are parametrised by an arbitrary function $f(z,\bar{z})$ on the sphere. $Q[f]$ can be written as the sum of a soft part ($Q_{\mathrm{soft}}[f]$) and a hard part ($Q_{\mathrm{hard}}[f]$). Upon asymptotic quantisation of the charge one can see that $Q_{\mathrm{soft}}[f]$ is responsible for the creation or annihilation of a soft graviton mode. Additionally one uses CK condition\footnote{The readers can refer to \cite{strom} for further details} to relate the positive helicity soft graviton mode with the negative helicity soft graviton mode. Finally the soft charge can be written as:
\beqa\label{gsoft}
Q_{\mathrm{soft}}[f]=\lim_{E_{p} \rightarrow 0} \ \frac{E_{p}}{4\pi\kappa}\int d^{2}w \  D^{2}_{w} f(w,\bar{w}) \ \Big(a_{-}(E_{p}~\hat{p}) + a_{+}^{\dagger}(E_{p}~\hat{p}) \Big),\nonumber \\
= \lim_{E_{p} \rightarrow 0} \ \frac{E_{p}}{4\pi\kappa} \int d^{2}w \  D^{2}_{\bar{w}}f(w,\bar{w}) \ \Big(a_{+}(E_{p}~\hat{p})+a_{-}^{\dagger}(E_{p}~\hat{p})\Big).
\eeqa
Here $\hat{p}$ denotes the direction of the soft graviton labelled by the coordinates $(w,\bar{w})$ and $\kappa^{2}=32\pi G$. $D_{w}$/$D_{\bar{w}}$ refers to the covariant derivative w.r.t to the $2$-sphere \\
%Using crossing symmetry of the scattering amplitude, one can rewrite the soft charges so as to include only one of the polarisations. i.e 
%\beqa\label{gsoftc}
%Q_{\mathrm{soft}}[f]=\lim_{E_{p} \rightarrow 0} \ \frac{E_{p}}{2\pi\kappa}\int d^{2}w \  D^{2}_{w} f(w,\bar{w}) \ a_{-}(E_{p}~\hat{p}), \nonumber \\
%= \lim_{E_{p} \rightarrow 0} \ \frac{E_{p}}{2\pi\kappa} \int d^{2}w \  D^{2}_{\bar{w}}f(w,\bar{w}) \ a_{+}(E_{p}~\hat{p}).
%\eeqa

The hard charge $Q_{\mathrm{hard}}[f]$ receives contribution from two terms. One of them includes the contribution of the energy momentum tensor of massless particles at null infinity, while the other includes the contribution of massive particles that will reach time-like infinity. The hard charge action on a massless and massive particle\footnote{The seminal work for outgoing/incoming particles being massive was done in \cite{alok:massive} .} can be written as:
\begin{equation}\label{ghard}
\begin{split}
Q_{\mathrm{hard}}[f]\ket{\mathrm{in}} &=\sum\limits_{i=\mathrm{in}}E_{k_{i}} \ f(\hat{k_{i}})\ket{\mathrm{in}}~~~ ;~~~Q_{\mathrm{hard}}[f]\ket{\mathrm{in}}_{\mathrm{massive}} =\sum\limits_{i=\mathrm{in}}m_{i} \ f_{H}(\vec{k_{i}}/m_{i})\ket{\mathrm{in}}_{\mathrm{massive}},
\\ 
\bra{\mathrm{out}}Q_{\mathrm{hard}}[f] &=\sum\limits_{i=\mathrm{out}}E_{k_{i}} \ f(\hat{k_{i}})\bra{\mathrm{out}}~~;~~_{\mathrm{massive}}\bra{\mathrm{out}}Q_{\mathrm{hard}}[f] =\sum\limits_{i=\mathrm{out}}m_{i} \ f_{H}(\vec{k_{i}}/m_{i})\bra{\mathrm{out}}_{\mathrm{massive}}.
\end{split}
\end{equation}
Here, the sum $\sum\limits_{\mathrm{in}}$ and $\sum\limits_{\mathrm{out}}$ is over all the hard particles in the ``in" and ``out" states. Energy and mass of the hard particles is denoted by $E_{k_{i}}$ and $m_{i}$ respectively. The unit spatial vector and $3-$momentum of the $i{\mathrm{th}}$ particle are denoted by $\hat{k_{i}}$ and $\vec{k_{i}}$ respectively. For a massive particle, $f_{H}(\vec{k_{i}}/m_{i})$ is defined by:
\begin{align}\label{fh}
f_{H}(\vec{k_{i}}/m_{i})=\int d^{2}w~ \mathcal{G}(\vec{k_{i}}/m_{i};w,\bar{w})f(w,\bar{w}).
\end{align}
where
\begin{align}
\mathcal{G}(\vec{k_{i}}/m_{i};w,\bar{w})=-\frac{1}{2\pi}D_{\bar{w}}^{2}\frac{\Big((\epsilon^{+}(w,\bar{w}))\cdot(k_{i}/m_{i})\Big)^{2}}{(p/E_{p})\cdot (k_{i}/m_{i})}.
\end{align}
Here $p\equiv E_{p}(1,\hat{p})$ denotes the $4$-momentum and $\epsilon^{+}(w,\bar{w})$ denotes the polarisation of the soft graviton labelled by the coordinates $(w,\bar{w})$.  The polarisation vector  $\epsilon^{+}(w,\bar{w})$ can be written as $\epsilon^{+}(w,\bar{w})=1/\sqrt{2}(\bar{w},1,-i,-\bar{w})$.
%Similarly the hard charge action on a massive particle can be written as
%\begin{equation}\label{ghard}
%\begin{split}
%Q_{f}^{\mathrm{hard}}\ket{\mathrm{in}} &=\sum\limits_{\mathrm{in}}E_{i} \ f_{H}(\hat{k_{i}}/m)\ket{\mathrm{in}} 
%\\ 
%\bra{\mathrm{out}}Q_{f}^{\mathrm{hard}} &=\sum\limits_{\mathrm{out}}E_{i} \ f_{H}(\hat{k_{i}})\bra{\mathrm{out}}
%\end{split}
%\end{equation}
\\
The Ward identity for supertranslation can be written as:
\beq\label{gward}
\bra{\mathrm{out}}[Q[f],\sa]\ket{\mathrm{in}}=0 \Leftrightarrow \bra{\mathrm{out}}[Q_{\mathrm{soft}}[f],\sa]\ket{\mathrm{in}}=-\bra{\mathrm{out}}[Q_{\mathrm{hard}}[f],\sa]\ket{\mathrm{in}}.
\eeq
For simplicity let us consider the ``in" and ``out" states to contain only massive particles. Then the Ward identity becomes
\begin{align}\label{pgward0}\nonumber
\lim_{E_{p} \rightarrow 0}\frac{E_{p}}{4\pi} \int d^{2}w D^{2}_{\bar{w}}f(w,\bar{w})& \Big(\bra{\mathrm{out}}a_{+}(E_{p}~\hat{p}) \sa\ket{\mathrm{in}}+\bra{\mathrm{out}}\sa~ a_{-}^{\dagger}(E_{p}~\hat{p})\ket{\mathrm{in}}\Big)
\\&=-\kappa\bigg[\sum\limits_{\mathrm{out}}m_{i} f_{H}(\vec{k_{i}}/m_{i})-\sum\limits_{\mathrm{in}}m_{i} f_{H}(\vec{k_{i}}/m_{i})\bigg]\sca.
\end{align}
Using crossing symmetry the above equation can be written as
\begin{align}\label{pgward}\nonumber
\lim_{E_{p} \rightarrow 0}\frac{E_{p}}{2\pi} \int d^{2}w D^{2}_{\bar{w}}f(w,\bar{w})& \Big(\bra{\mathrm{out}}a_{+}(E_{p}~\hat{p}) \sa\ket{\mathrm{in}}\Big)
\\&=-\kappa\bigg[\sum\limits_{\mathrm{out}}m_{i} f_{H}(\vec{k_{i}}/m_{i})-\sum\limits_{\mathrm{in}}m_{i} f_{H}(\vec{k_{i}}/m_{i})\bigg]\sca.
\end{align}
\\
The Weinberg soft graviton theorem \cite{weinberg} is given by
\beq\label{weinberg}
\lim_{E_{p} \rightarrow 0}E_{p} \bra{\mathrm{out} }a_{+} (E_{p}~\hat{p}) \  \sa \ket{\mathrm{in}}=\frac{\kappa}{2}\Big(\sum \limits_{i=\mathrm{out}}\frac{(\epsilon^{+}(w,\bar{w}) \cdot k_{i})^{2}} {(p/E_{p})  \cdot k_{i}}-\sum \limits_{i=\mathrm{in}}\frac{(\epsilon^{+}(w,\bar{w}) \cdot k_{i})^{2}} {(p/E_{p})  \cdot k_{i}}\Big)\bra{\mathrm{out}}\sa\ket{\mathrm{in}}.
\eeq
%where the soft graviton direction is parametrized by ($w,\bar{w}$) and its polarization is given by $\epsilon^{+}(w,\bar{w})=1/\sqrt{2}(\bar{w},1,-i,-\bar{w})$. 
We adopt the notation,
\beq\label{lfactor}
S^{(0)}(\hat{p};k_{i})\equiv \frac{(\epsilon^{+}(w,\bar{w}) \cdot k_{i})^{2}} {(p/E_{p})  \cdot k_{i}}.
\eeq 
with which, the leading soft factor in the r.h.s. of \eqref{weinberg} can be written as:
\begin{align}\label{crazy-notation}
S^{(0)}(\hat{p};\{k_{i}\})&\equiv \sum \limits_{i=\mathrm{out}}\frac{(\epsilon^{+}(w,\bar{w}) \cdot k_{i})^{2}} {(p/E_{p})  \cdot k_{i}}-\sum \limits_{i=\mathrm{in}}\frac{(\epsilon^{+}(w,\bar{w}) \cdot k_{i})^{2}} {(p/E_{p})  \cdot k_{i}},\\
&\equiv \sum \limits_{i=\mathrm{out}}S^{(0)}(\hat{p};k_{i})-\sum \limits_{i=\mathrm{in}}S^{(0)}(\hat{p};k_{i}).
\end{align}
%\beq
%\sum \limits_{i}\frac{(\epsilon^{+}(w,\bar{w}) \cdot k_{i})^{2}} {(p/E_{p})  \cdot k_{i}}\equiv \equiv \sum \limits_{i}S^{(0)}(p;k_{i})
%\eeq
Now one can show the equivalence of the Ward identity with  soft theorem by choosing a particular $f(z,\bar{z})$ namely,
\begin{align}\label{leadfn}
f(z,\bar{z}) = s(z,\bar{z};w,\bar{w}) \equiv \frac{1+w\bar{w}}{1+z\bar{z}} \cdot \frac{\bar{w}-\bar{z}}{w-z}.
\end{align}

In the rest of the paper, for simplicity of the calculations we choose the specific form of $f(z,\bar{z})$ in \eqref{leadfn} which led us to the soft theorem. After choosing this, we denote the soft charge and hard charge as
$Q_{\mathrm{soft}}(\hat{p})$ and $Q_{\mathrm{hard}}(\hat{p})$ respectively. The soft charge therefore becomes
\begin{align}\label{gscharge1}
Q_{\mathrm{soft}}(\hat{p})=\lim\limits_{E_{p}\rightarrow 0}~\frac{1}{2}E_{p}\Big(a_{+}(E_{p}~\hat{p})+a_{-}^{\dagger}(E_{p}~\hat{p})\Big).
\end{align}
In this paper we are always concerned with the action of the soft operator at the level of scattering amplitudes. Therefore using the notion of crossing symmetry we can always relate an outgoing positive helicity soft graviton with a negative helicity soft graviton. Hence \ref{gscharge1} can be further written as
\begin{align}\label{gscharge2}
Q_{\mathrm{soft}}(\hat{p})=\lim\limits_{E_{p}\rightarrow 0}E_{p}a_{+}(E_{p}~\hat{p}).
\end{align} 
The action of hard charge can be written as
\begin{align}\label{ghcharge2}
Q_{\mathrm{hard}}(\hat{p})\ket{k}=-\frac{\kappa}{2}S^{(0)}(\hat{p};k)\ket{k}.
\end{align}
where $S^{(0)}(\hat{p};k)$ is already defined in \eqref{lfactor}.

\subsection{Dressed States in Gravity}\label{gr2}
Following \cite{strom-fadeev}, a number of recent works \cite{akhoury2,akhoury3,porrati1,porrati2,porrati3,porrati4,gomez,sever,hirai} have analysed the intricate relationship between the spontaneous breaking of asymptotic symmetries, the corresponding existence of soft modes as Goldstone modes and asymptotic Hilbert space which comprises of coherent states of such soft modes. A number of these works (notably \cite{akhoury2,akhoury3,sever,gomez}) have argued that under certain conditions which we summarise below, the states in which asymptotic conservation laws are satisfied as Ward identities are precisely the well known dressed states in which S-matrix is infrared finite in the case of QED and gravity\cite{akhoury1}\cite{fk}. 

We revisit the earlier analysis \cite{akhoury2,akhoury3} below in the context of perturbative gravity but with an eye towards QCD. More in detail, we attempt to recast some of the main ideas in \cite{akhoury2,akhoury3} by asking the following question: if we assume that supertranslations are a symmetry of the perturbative S-matrix, what are the additional assumptions we need in order to derive the precise form of dressed states in perturbative gravity? 

In the next section, we review and extract the key ideas contained in \cite{akhoury2,akhoury3} which will help us determine the asymptotic Hilbert space from the existence of asymptotic symmetries in QCD. 
%One of the seminal attempts to construct an infra-red finite  The notable work being the construction of Fadeev and Kullish in QED (cite paper). Their work involved ``dressing" the charged massive Fock states by a coherent cloud of low energy photons. The S-matrix between such states were shown to be IR finite. This was later extended to perturbative gravity in \cite{akhoury1}, in which the states are dressed by a coherent cloud of low energy gravitons. 
%Such a dressed state can be written as
%\begin{align}
%e^{R(k)}\ket{k}=\exp{\Big(\kappa \int~ \widetilde{d^3q}~\frac{k^{\mu}k^{\nu}} {q  \cdot k}(a^{\dagger}_{\mu \nu}(q)-a_{\mu \nu}(q)\Big)}\ket{k}
%\end{align}
%where $a^{\dagger}_{\mu \nu}(q)$ and $a_{\mu \nu}(q)$ are the graviton creation and annihilation operators the integral over $q$ runs only over low energy gravitons.
%In a later work, \cite{akhoury2,akhoury3} they studied the implications of BMS supertranslation on S-matrix involving dressed states. In \cite{akhoury2}, it was shown that BMS supertranslation charge gives rise to infinite set of degenerate vaccuua, labelled by soft graviton modes. The scattering amplitude between the ``dressed" states built from one vaccua to another degenerate vaccua is zero. This is what we call as the orthogonality condition.\\

\subsection{Faddeev-Kulish States in Gravity from BMS Supertranslations} \label{gr3}

In \cite{akhoury3}, the authors showed that conservation of BMS charge leads to asymptotic states in which S-matrix elements are infra-red finite\footnote{The analogous analysis in QED was done in \cite{sever,strom-fadeev}.}. The argument can be summarised as follows. 
\\\\
Let us consider all the degenerate vacua as eigenstates of the BMS supertranslation soft charge and consider a transition between scattering states built over such vaccua. Let $\ket{\mathrm{N}}$ denotes the eigenstate of the soft operator $Q_{\mathrm{soft}}(\hat{p})$ (defined as in \eqref{gscharge1}). Let $\ket{\mathrm{N_{out}},\mathrm{out}}$/$\ket{\mathrm{N_{in}},\mathrm{in}}$ denotes the outgoing/incoming states which are the eigenstates of $Q_{\mathrm{soft}}(\hat{p})$ with eigenvalues $\mathrm{N_{out}}(\hat{p})$/$\mathrm{N_{in}}(\hat{p})$. i.e 
%\textcolor{red}{why is the symbol marked in red with a sub-script and all other soft charges with super-script? Also what happened to $f$? How is the following equation meaningful without $f$. Is it for a fixed $f$ or for all $f$s.}
\begin{align}
Q_{\mathrm{soft}}(\hat{p})\ket{\mathrm{N_{out}},\mathrm{out}}=\mathrm{N_{out}}(\hat{p})\ket{\mathrm{N_{out}},\mathrm{out}} ~~~,~~~ Q_{\mathrm{soft}}(\hat{p})\ket{\mathrm{N_{in}},\mathrm{in}}=\mathrm{N_{in}}(\hat{p})\ket{\mathrm{N_{in}},\mathrm{in}}.
\end{align}
The supertranslation Ward identity between such states can be written as:
\begin{align}\label{clg3}
\bra{\mathrm{N_{out}},\mathrm{out}}[Q_{\mathrm{soft}}(\hat{p}),\sa]\ket{\mathrm{N_{in}},\mathrm{in}}=-\bra{\mathrm{N_{out}},\mathrm{out}}[Q_{\mathrm{hard}}(\hat{p}),\sa]\ket{\mathrm{N_{in}},\mathrm{in}}.
\end{align} 
The above expression can be evaluated as:
\begin{align}
\Big(\mathrm{N_{out}}(\hat{p}) - \mathrm{N_{in}}(\hat{p}) \Big)\bra{\mathrm{N_{out}},\mathrm{out}}\sa\ket{\mathrm{N_{in}},\mathrm{in}}=\Omega_{\mathrm{soft}}(\hat{p})\bra{\mathrm{N_{out}},\mathrm{out}}\sa\ket{\mathrm{N_{in}},\mathrm{in}}.
\end{align}
where $\Omega_{\mathrm{soft}}(\hat{p})$ is given by the soft factor \eqref{crazy-notation}. In evaluating the r.h.s of \eqref{clg3} one assumes that the hard charge has a trivial action on the degenerate vaccua.
Now the above equation suggests two possibilities:
\begin{align}\label{cg}
\mathrm{N_{out}}(\hat{p}) - \mathrm{N_{in}}(\hat{p}) - \Omega_{\mathrm{soft}}(\hat{p}) =0.
\end{align}
or
\begin{align}
\bra{\mathrm{N_{out}},\mathrm{out}}\sa\ket{\mathrm{N_{in}},\mathrm{in}}=0.
\end{align}
Now if we demand that the transition amplitude between such degenerate vaccua is non-trivial then,
\begin{align}\label{fkconstraint}
\mathrm{N_{out}}(\hat{p}) - \mathrm{N_{in}}(\hat{p})  =\Omega_{\mathrm{soft}}(\hat{p}).
\end{align}
The authors in \cite{akhoury3} proposed an ansatz for the dressing operator for constructing such states which can be written as
\begin{align}
e^{R_{N}}= \exp{\Big(\kappa \int d[k]~\rho (k)\int^{\Lambda}~ d[q]~N^{\mu\nu}(q;k)(a^{\dagger}_{\mu \nu}(q)-a_{\mu \nu}(q)\Big)}.
\end{align}
We will study this operator in detail in the next section. They further showed that the Faddeev-Kulish states belong to such class of dressed states and these satisfy the contraint. \eqref{fkconstraint}.\\ 
%\textcolor{red}{WRITE INTRODUCTION PROPERLY
%	Recently there has been series of works relating the asymptotic symmetries and Fadeev Kulish states in QED and gravity. It has been shown in QED that the equivalence of soft theorem with the Ward Identity is outcome of not working with asymptotic states\cite{sever}. This was later extended to gravity \cite{akhoury2}. 
%	In (cite Akourys's work) it was shown that the asymptotic states can be parametrized by the superselection sector to which they belong. One yields a IR finite S-matrix if one considers a transition between states belonging to same supertranslation sector. They further showed that BMS supertranslation charge creates a infinite set of degenerate vaccua. Therefore it is natural to characterize the vaccum state by the eigen states of the supertranslation soft charge. If one considers the Supertranslation Ward Identity between states built from these vaccua , one can arrive at set of conservation laws which gives us the condition for getting an IR finite S-matrix. The scattering amplitude construced from such states is exactly equal to the Fadeev Kulish amplitudes.
%}
\subsection{A Closer Look at the Derivation of Dressed States from Symmetries}\label{gr4}
%In this section we revisit the analysis of  \cite{akhoury2}\cite{akhoury3}. Our goal is to generalise these ideas to obtain Asymptotic states in QCD. In order to do so, we encounter certain subtleties in the derivation of dressed states in gravity from ST Ward identities which we attempt to address below.
As we recalled above, the derivation of dressed states consistent with the supertranslation conservation law relied on three key inputs. 
\begin{itemize}
	\item The dressed state is an eigenstate of the soft charge, 
	\item The hard charge has a trivial action on the dressing (as it had no gravitational contribution) and 
	\item S-matrix elements evaluated in the dressed states are non-trivial.
\end{itemize} 
However there are some caveats here that need to be emphasised from our perspective.
\begin{enumerate}
	\item 	As we review below, due to the fact that vacuum is shifted by supertranslation soft charge, the dressed state for any choice of dressing can not be an eigenstate of the soft charge. Due to this it may appear that the analysis presented in the previous section \ref{gr2} is inconsistent. However as we show below, this analysis can be made consistent if we demand that the dressed states satisfy a constraint called orthogonality condition.\footnote{The known construction of dressed states such as Faddeev-Kulish states do satisfy this condition as was shown  in \cite{sever,akhoury3}.} Analysis of the orthogonality condition will be central to us in the derivation of QCD asymptotic states. \label{caveat1}
	\item  If we work in linearised gravity where the hard charge has no contribution from gravitational news tensor, it is indeed true that the hard charge commutes with the dressing. However if the supertranslation hard charge contains contribution from gravitational field one needs to be careful with the commutators of the hard charge with the dressing. We will address this issue below by using the known action of supertranslation charge on the Goldstone mode conjugate to the soft mode \cite{strom}.  
	\label{caveat2}
\end{enumerate}
%\noindent{\bf (1)}	As we show below, due to the fact that vacuum is shifted by supertranslation soft charge, the dressed state for any choice of dressing can not be an eigenstate of the soft charge. Due to this it may appear that the analysis presented in the previous section \ref{fkg} is inconsistent. However as was shown in \cite{sever,akhoury3}, for known dressings like Faddeev-Kulish dressings, this shifted state is ``null" in the sense that it is orthogonal to all the states in the dressed Hilbert space. And this makes the analysis of \cite{akhoury3} self-consistent. But as we are trying to derive the nature of asymptotic states from the conservation laws associated to asymptotic symmetries, we can not take recourse to known dressed states like Faddeev-Kulish states. This point will be central to us in the derivation of QCD asymptotic states. 
% \\\\
% \noindent{\bf (2)} If we work in linearised gravity where the hard charge has no contribution from gravitational News tensor, then it is indeed true that the hard charge commutes with the dressing. However if the supertranslation hard charge contains contribution from gravitational field then this no longer remains true. \\\\
These caveats turn out to be even more severe in QCD and hence in order apply the analysis of \cite{strom-fadeev,akhoury2,akhoury3} in that case, we determine the relationship between dressed states and asymptotic conservation laws in a slightly different manner, such that both the caveats mentioned above become explicit.
%Let us assume that the dressed states are constructed by the action of a dressing operator (denoted by $e^{R_{N}}$ where $N$ is an arbitrary function) on a ``bare" state. The dressed state therefore can be written as
%\begin{align}
%e^{R_{N}}\ket{k}=e^{R_{N}(k)}\ket{k}
%\end{align}
Namely, our goal is to explore to what extent the infinity of supertranslation conservation laws can constrain the form of the dressed states. To address this question, we start with four assumptions.
\begin{enumerate}
	\item Supertranslation symmetry is a symmetry of the quantum S-matrix.\label{st1}
	\item Soft graviton modes satisfy what we call orthogonality relations with respect to the asymptotic states of the theory. Orthogonality condition simply means that if we consider a state $\vert\psi\rangle$ which is a tensor product of (dressed) finite energy state and one or more soft graviton state, then this state $\vert\psi\rangle$ is orthogonal to all the (dressed) finite energy states.\label{st2}
	\item The S-matrix elements are non-trivial. \label{st3} 
	\item The supertranslation hard charge has a trivial action on the dressing,\label{st41}	
\end{enumerate}
%The assumption \ref{st41} is motivated by looking at the action of super-translation on soft modes as described in \cite{strom}. 
%
%We give a detailed computation justifying our assumption in appendix \ref{2d}. 
As we show below, by using all the assumptions mentioned above one can determine the dressed states in gravity.
% Later on in the calculations we show a heuristic argument  the assumption \eqref{st41}. 
In section \ref{gr6} we show that if one follows along the lines of \cite{strom} one obtains a different type of dressing. It turns out that in this analysis one need not use assumption \eqref{st41}. 

To determine the dressed states which respect the above mentioned assumptions, we start with an ansatz for the dressing operator following	\cite{strom-fadeev,akhoury1,akhoury2}	
%Following  we start with the following ansatz for the dressing operator 
\begin{align}\label{dresso}
e^{R_{N}}= \exp{\Big(\kappa \int d[k]~\rho (k)\int^{\Lambda}~ d[q]~N^{\mu\nu}(q;k)(a^{\dagger}_{\mu \nu}(q)-a_{\mu \nu}(q)\Big)}.
\end{align}

where $\rho (k)\equiv b^{\dagger}(k)b(k)$ is the number operator for the external massive particles (we are considering massive scalar particles for simplicity and $b^{\dagger}(k)$, $b(k)$ are the creation and annihilation operators for scalar particle respectively), $\kappa^{2}=32\pi G$ and $N^{\mu\nu}(q;k)$ is an arbitrary real function which has a pole in $E_{q}$. Here $d[k]\equiv \frac{d^{3}k}{(2\pi)^{2}2E_{k}}$ is the Lorentz invariant measure. $\Lambda$ is an upper cut off for the integral over $q$ to ensure that the dressing comprises only of low energy gravitons.
%\textcolor{red}{I think you should remove $\Lambda$ from the cut-off here and argue that it needs to be there from the perspective of hard charge commutation with dressing.}
$a^{\dagger}_{\mu \nu}(q)$ and $a_{\mu \nu}(q)$ are the graviton creation and annihilation operators respectively which can be written in the polarisation basis as
\begin{align}\label{polbasis}
a^{\dagger}_{\mu \nu}(q)=\sum_{r=\pm}\epsilon^{r}_{\mu\nu}(q)a^{r\dagger}(q),~~~a_{\mu \nu}(q)=\sum_{r=\pm}\epsilon^{*r}_{\mu\nu}(q)a^{r}(q).
\end{align}
 These operators obey the commutation relation
\begin{align}\label{acomm}
[a^{r}(q),a^{s\dagger}(q')]=\delta^{rs}(2E_{q})(2\pi)^{3}\delta^{3}(\vec{q}-\vec{q}~').
\end{align}
%where
%\begin{align}
%I_{\mu\nu\rho\sigma}\equiv \eta_{\mu\rho}\eta_{\nu\sigma}+\eta_{\mu\sigma}\eta_{\nu\rho}-\eta_{\mu\nu}\eta_{\rho\sigma}
%\end{align}
%It is also easy to see that the asymptotic operator \eqref{dresso} is anti-Hermitian.

A dressed state is constructed using the action of the asymptotic operator \eqref{dresso} on a ``bare" state. Here we consider the dressing on a massive scalar field defined by
\begin{align}
\varphi(x)=\int d[k] \Big[b(k)e^{i k.x}+ b^{\dagger}(k)e^{-i k.x}\Big].
\end{align}
The creation and annihilation operators of the scalar particle obey the commutation relation
\begin{align}
[~b(k),b^{\dagger}(k')~]=(2\pi)^{3}(2E_{k})\delta^{3}(\vec{k}-\vec{k}~').
\end{align}
\\
The dressed scalar state is then defined by
\begin{align}\label{dress}
e^{R_{N}}\ket{k}=e^{R_{N}}b^{\dagger}(k)\ket{0}=[e^{R_{N}},b^{\dagger}(k)]\ket{0}= e^{R_{N}(k)}b^{\dagger}(k)\ket{0}.
\end{align}
where
\begin{align}\label{gansatz}
e^{R_{N}(k)}=\exp{\Big(\kappa \int^{\Lambda} d[q]~N^{\mu\nu}(q;k)(a^{\dagger}_{\mu \nu}(q)-a_{\mu \nu}(q)\Big)}.
\end{align}
The action of the asymptotic operator on a multiparticle states can also be found similarly,
\begin{align}\label{mpd}
e^{R_{N}}\ket{k_{1},k_{2}\ldots k_{n}}=e^{R_{N}(k_{1})}e^{R_{N}(k_{2})}\ldots e^{R_{N}(k_{n})}\ket{k_{1},k_{2}\ldots k_{n}}.
\end{align}
where $\ket{k_{1},k_{2},\ldots, k_{n}}$ denotes a multiparticle state generated by the action of creation operators $b^{\dagger}(k_{1}),$ $b^{\dagger}(k_{2}),\ldots ,b^{\dagger}(k_{n})$ on the Fock vaccum. From \eqref{mpd} it is clear that the dressing operator factorises in the ``hard" particle space.
\\\\
Having defined the dressed state we demonstrate caveat \eqref{caveat1} which we discussed in the beginning of this section. Let us consider the action of the soft operator $Q_{\mathrm{soft}}(\hat{p})$ \eqref{gscharge1} on a dressed state. The soft charge can be written as
%For simplicity let us assume the specific form of $f(z,\bar{z})$ which led us to the equivalence of the supertranslation Ward Identity with the leading soft graviton theorem in the undressed state \eqref{leadfn}.\textcolor{red}{As we discussed over phone, this needs to be re-written depending on when one introduces $Q^{\mathrm{soft}}$ from $Q[f]$.} After choosing this function we denote the hard and soft charges as $\qhard$ and $\qsoft$ respectively. The soft charge can be therefore written as
\beqa\label{leading-soft1}
Q_{\mathrm{soft}}(\hat{p})
= \lim_{E_{p} \rightarrow 0} \ \frac{E_{p}}{2} \Big(a_{+}(E_{p}~ \hat{p})+a_{-}^{\dagger}(E_{p}~ \hat{p})\Big).
\eeqa
%where $\hat{x_{p}}$ corresponds to the direction of the soft graviton.
The action of $Q_{\mathrm{soft}}(\hat{p})$ on the dressed state defined by \eqref{dress} can be expressed as
\begin{align}\label{softaction}
Q_{\mathrm{soft}}(\hat{p})e^{R_{N}}\ket{k}=[Q_{\mathrm{soft}}(\hat{p}),e^{R_{N}(k)}]\ket{k} +e^{R_{N}(k)} Q_{\mathrm{soft}}(\hat{p})\ket{k}.
\end{align}
Let us first compute the commutator term
\begin{align}\label{sdressc}\nonumber
[Q_{\mathrm{soft}}(\hat{p}),e^{R_{N}(k)}]&=[Q_{\mathrm{soft}}(\hat{p}),R_{N}(k)]~e^{R_{N}(k)},\\\nonumber
&=\lim_{E_{p} \rightarrow 0} \ \frac{E_{p}}{2} \Big[a_{+}(E_{p}~ \hat{p})+a_{-}^{\dagger}(E_{p}~ \hat{p}),\kappa \int^{\Lambda} d[q]~N^{\mu\nu}(q;k)(a^{\dagger}_{\mu \nu}(q)-a_{\mu \nu}(q)\Big]e^{R_{N}(k)},\\
&=\lim\limits_{E_{p}\rightarrow 0}~\kappa E_{p} N^{\mu\nu}(p;k)\epsilon_{\mu\nu}^{+} (\hat{p})e^{R_{N}(k)}.
\end{align}
where we used \eqref{acomm}and \eqref{polbasis} to compute the last line from the second line.\\
Therefore \eqref{softaction} can be written as
\begin{align}\label{eigen}
Q_{\mathrm{soft}}(\hat{p})e^{R_{N}}\ket{k}=\lim\limits_{E_{p}\rightarrow 0}~\kappa E_{p} N^{\mu\nu}(p;k)\epsilon_{\mu\nu}^{+} (\hat{p})e^{R_{N}(k)}\ket{k} + e^{R_{N}(k)}Q_{\mathrm{soft}}(\hat{p})\ket{k}.
\end{align}
The second term in the above expression can be written as
\begin{align}\label{extra}
e^{R_{N}(k)}Q_{\mathrm{soft}}(\hat{p})\ket{k}=e^{R_{N}(k)}b^{\dagger}(k)Q_{\mathrm{soft}}(\hat{p})\ket{0}.
\end{align}
 This corresponds to the action of the asymptotic operator on a one particle state built from a supertranslated vacuum.\footnote{One can also use the prescription \eqref{gscharge2} for $Q_{\mathrm{soft}}(\hat{p})$ and claim that the extra term \eqref{extra} vanishes. But this subtlety will again arise when one considers its action on a outgoing dressed state (bra).}
If this term vanishes then clearly the dressed state would be the eigenstate of the soft operator. But as the supertranslation charge shifts vacuum instead of annihilating it, this is not true. 
\\
In order to determine the dressed states in gravity let us start with our assumption \eqref{st1} that supertranslation is a symmetry of the S-matrix. We then have the corresponding Ward identity
\begin{align}\label{gward1}
{}_{\mathrm{d}}\bra{\mathrm{out}}[Q(\hat{p}),\mathrm{S}]\ket{\mathrm{in}}_{\mathrm{d}}=0.
\end{align}
where $Q(\hat{p})=Q_{\mathrm{soft}}(\hat{p}) + Q_{\mathrm{hard}}(\hat{p})$ (defined in \eqref{gscharge1} and \eqref{ghcharge2}).
${}_{\mathrm{d}}$$\bra{\mathrm{out}}$, 
$\ket{\mathrm{in}}_{\mathrm{d}}$ denotes the dressed ``out" and ``in" states, which are given as
\begin{align}\nonumber
{}_{\mathrm{d}}\bra{\mathrm{out}}&=\braout\odress= \bra{\mathrm{out}}\exp{\Big(-\kappa \sum_{i =\mathrm{out}}\int^{\Lambda} d[q]~ N_{i}^{\mu\nu}(q;k_{i})(a^{\dagger}_{\mu \nu}(q)-a_{\mu \nu}(q)\Big)},\\
\ket{\mathrm{in}}_{\mathrm{d}}&= \idress\ketin=\exp{\Big(\kappa \sum_{i =\mathrm{in}} \int^{\Lambda} d[q]N_{i}^{\mu\nu}(q;k_{i})(a^{\dagger}_{\mu \nu}(q)-a_{\mu \nu}(q)\Big)}\ket{\mathrm{in}}.
\end{align}
\\
And $k_{i}$ denotes the momentum of the external particles in the outgoing and incoming states.
\\
%The Ward Identity can then be written as
%\begin{align}\label{gward1}
%\braout\odress[Q_{f},S]\idress\ketin=0
%\end{align}
%\\
After expressing the charge as the sum of soft and hard part, the Ward identity \eqref{gward1} becomes
\begin{align}\label{gward2}
\braout\odress[Q_{\mathrm{soft}}(\hat{p}),S]\idress\ketin=-\braout\odress[Q_{\mathrm{hard}}(\hat{p}),S]\idress\ketin.
\end{align}
\\
Let us now consider the l.h.s of \eqref{gward2}. This can be expanded as,
\begin{align}\label{gwardsoft}\nonumber
\braout Q_{\mathrm{soft}}(\hat{p})\odress\sa&\idress\ketin-\braout\odress\sa\idress Q_{\mathrm{soft}}(\hat{p})\ketin\\
&+\braout[\odress,Q_{\mathrm{soft}}(\hat{p})]\sa\idress\ketin-\braout\odress\sa[Q_{\mathrm{soft}}(\hat{p}),\idress]\ketin.
\end{align}
\\
We can now use assumption \eqref{st2}, namely the orthogonality condition which implies that both of the first two terms of the above equation vanish. Therefore we are left with,
\begin{align}\label{gwardsoft2}
\braout[\odress,Q_{\mathrm{soft}}(\hat{p})]\sa\idress\ketin-\braout\odress\sa[Q_{\mathrm{soft}}(\hat{p}),\idress]\ketin .
\end{align}
\\
The commutators in the above expression can be evaluated using \eqref{sdressc} as,
\begin{align}\label{cloudc}
[\odress,Q_{\mathrm{soft}}(\hat{p})]=\lim\limits_{E_{p}\rightarrow 0}~\kappa E_{p} N_{\mathrm{out}}(p)\odress~~~,
~~~~~[Q_{\mathrm{soft}}(\hat{p}),\idress]=\lim\limits_{E_{p}\rightarrow 0}~\kappa E_{p} N_{\mathrm{in}}(p)\idress.
\end{align} 
where
\begin{align}\label{noutnin}
N_{\mathrm{out}}(p)=\sum_{i =\mathrm{out}}N^{\mu\nu}_{i}(p;k_{i})\epsilon_{\mu\nu}^{+}(\hat{p})~~~,~~~~
N_{\mathrm{in}}(p)=\sum_{i =\mathrm{in}}N^{\mu\nu}_{i}(p;k_{i})\epsilon_{\mu\nu}^{+}(\hat{p}).
\end{align}
\\
Using these we can evalute l.h.s of \eqref{gward2} to
\begin{align}\label{gwsoft}\nonumber
\braout[\odress,Q_{\mathrm{soft}}(\hat{p})]\sa\idress\ketin-&\braout\odress\sa[Q_{\mathrm{soft}}(\hat{p}),\idress]\ketin,\\ &=\kappa \lim\limits_{E_{p}\rightarrow 0}E_{p}\Big(N_{\mathrm{out}}(p)-N_{\mathrm{in}}(p)\Big)\braout\odress\sa\idress\ketin.
\end{align}
\\
The r.h.s of \eqref{gward2} can be expanded as
\begin{align}\nonumber
\braout Q_{\mathrm{hard}}(\hat{p})&\odress\sa\idress\ketin-\braout\odress\sa\idress Q_{\mathrm{hard}}(\hat{p})\ketin,\\
&+\braout[\odress,Q_{\mathrm{hard}}(\hat{p})]\sa\idress\ketin-\braout\odress\sa[Q_{\mathrm{hard}}(\hat{p}),\idress]\ketin.
\end{align}
\\
Using the action of the hard charges the first two terms in the above expression gives
\begin{align}\label{gwhard}\nonumber
\braout Q_{\mathrm{hard}}(\hat{p})\odress\sa\idress\ketin-\braout&\odress\sa\idress Q_{\mathrm{hard}}(\hat{p})\ketin=
\\&-\frac{\kappa}{2}S^{(0)}(\hat{p};\{k_{i}\})\braout\odress\sa\idress\ketin.
\end{align}
where
\begin{align}
S^{(0)}(\hat{p};\{k_{i}\})=\sum \limits_{i=\mathrm{out}}\frac{(\epsilon^{+}(w,\bar{w}) \cdot k_{i})^{2}} {(p/E_{p})  \cdot k_{i}}-\sum \limits_{i=\mathrm{in}}\frac{(\epsilon^{+}(w,\bar{w}) \cdot k_{i})^{2}} {(p/E_{p} )  \cdot k_{i}}.
\end{align}
\\
Using the expressions \eqref{gwhard} and \eqref{gwsoft} the Ward identity \eqref{gward2} can finally be written as
\begin{align}\nonumber\label{clg}
\kappa\lim\limits_{E_{p}\rightarrow 0}E_{p}&\Big(N_{\mathrm{out}}(p)-N_{\mathrm{in}}(p)\Big)\braout\odress\sa\idress\ketin,\\\nonumber
&=\frac{\kappa}{2}S^{(0)}(\hat{p};\{k_{i}\})\braout\odress\sa\idress\ketin- \braout[\odress,Q_{\mathrm{hard}}(\hat{p})]\sa\idress\ketin\\&~~~~~~~~~~~~~~~~~~~~~~~~~~~~~~~~~~~~~~~~~~~~~~~~~~~~~+\braout\odress\sa[Q_{\mathrm{hard}}(\hat{p}),\idress]\ketin.
\end{align}
At this point let us recall caveat \eqref{caveat2}. If the hard charge has contribution from the gravitational news tensor then $Q_{\mathrm{hard}}(\hat{p})$ does not commute with the dressing and one will not retrieve the usual conservation laws \eqref{cg}. In appendix \eqref{gra2} we show that the contribution from the commutator of hard charge with the dressing is of $\mathcal{O}(\Lambda)$ where $\Lambda$ is the upper cut off in the dressing operator. If $\Lambda$ is sufficiently small then one could ignore these terms. Hence the last two terms in the r.h.s of \eqref{clg} vanishes by assumption \eqref{st41}. Therefore
\begin{align}\label{clg2}
\kappa\lim\limits_{E_{p}\rightarrow 0}E_{p}\Big(N_{\mathrm{out}}(p)-N_{\mathrm{in}}(p)\Big)\braout\odress\sa\idress\ketin,
=\frac{\kappa}{2}S^{(0)}(\hat{p};\{k_{i}\})\braout\odress\sa\idress\ketin.
\end{align}
We can now use assumption \eqref{st3} to show that, 
\begin{align}\label{sconstraint}
\lim\limits_{E_{p}\rightarrow 0}E_{p} \Big(N_{\mathrm{out}}(p)-N_{\mathrm{out}}(p)\Big)=\frac{1}{2}S^{(0)}(\hat{p};\{k_{i}\}).
\end{align}
or
\begin{align}\nonumber
\lim\limits_{E_{p}\rightarrow 0}E_{p} \Big(\sum_{i =\mathrm{out}}N^{\mu\nu}_{i}(p;k_{i})\epsilon_{\mu\nu}^{+}(\hat{p})-\sum_{i =\mathrm{in}}N^{\mu\nu}_{i}&(p;k_{i})\epsilon_{\mu\nu}^{+}(\hat{p})\Big)\\
&=\frac{1}{2}\Big( \sum_{i =\mathrm{out}}\frac{(\epsilon^{+}(w,\bar{w}) \cdot k_{i})^{2}} {(p/E_{p})  \cdot k_{i}}-\sum_{i =\mathrm{in}}\frac{(\epsilon^{+}(w,\bar{w}) \cdot k_{i})^{2}} {(p/E_{p})  \cdot k_{i}} \Big).
\end{align}
\\
From the above expression we can associate naturally,
\begin{align}\label{ldress}
N^{\mu\nu}_{i}(p;k_{i})=\frac{1}{2}\frac{k_{i}^{\mu}k_{i}^{\nu}} {p \cdot k_{i}}.
\end{align}
\\
Thus we have recovered the dressing factor for each external particle.
\\
Substituting \eqref{ldress} in the dressing operator \eqref{gansatz} we finally get the dressed state as
\begin{align}\label{lgdress}
e^{R_{N}(k)}\ket{k}=\exp{\Big(\frac{\kappa}{2} \int^{\Lambda}~ d[q]~\frac{k^{\mu}k^{\nu}} {q  \cdot k}(a^{\dagger}_{\mu \nu}(q)-a_{\mu \nu}(q)\Big)}\ket{k}.
\end{align}
\\
This matches with the Faddeev-Kulish states in gravity \cite{akhoury2}.
\subsection{Orthogonality Relations for Multiple Soft Graviton Insertions}\label{gr5}

Having derived the dressed states in gravity, we would now like to see whether these states decouple a finite number of soft graviton modes. i.e., we would like to see whether the orthogonality condition (assumption \eqref{st2}) holds for more than one soft graviton mode. This analysis becomes important in QCD as it leads to a modification of the dressing factor in QCD. We will explain this detail in the QCD section. Since we already know that at the leading level, the multiple soft graviton theorems are not independent in the sense that each of the soft factors can be determined using just the leading single soft factor, one does not expect any issue for the orthogonality condition involving multiple soft graviton modes. We will explicitly prove this in this section.
\\
Without loss of generality let us consider a case in which vacuum is shifted by two soft graviton modes. i.e., we would like to show 
\begin{align}\label{dog}
\braout Q_{\mathrm{soft}}(\hat{p}_{1})Q_{\mathrm{soft}}(\hat{p}_{2})\odress\sa\idress\ketin=0.
\end{align}
where $Q_{\mathrm{soft}}(\hat{p}_{1})$ and $Q_{\mathrm{soft}}(\hat{p}_{2})$ are  soft graviton modes given by
\begin{align}
Q_{\mathrm{soft}}(\hat{p}_{1})&=\lim\limits_{E_{p_{1}}\rightarrow 0}~~E_{{p_{1}}} ~a_{+}(E_{{p_{1}}}~\hat{p}_{1}),\\
Q_{\mathrm{soft}}(\hat{p}_{2})&=\lim\limits_{E_{{p_{2}}}\rightarrow 0}~~E_{p_{2}} ~a_{+}(E_{p_{2}}~\hat{p}_{2}).
\end{align}\\
and $e^{R_{N}}$ is the dressing operator already derived in previous section \eqref{lgdress}. 
The l.h.s of \eqref{dog} can be evaluated as done in \cite{akhoury2,sever}. It will receive contribution from three terms. 
\begin{enumerate}
	\item Both the soft gravitons are connected to  the external particles.\label{do1}
	\item One of the soft graviton is connected to the external particle and the other is connected to the dressing operator.\label{do2}
	\item Both the soft gravitons are connected to the dressing operator. \label{do3}
\end{enumerate}
The first contribution \eqref{do1} can be evaluated by the leading double soft graviton theorem\footnote{One can use either consecutive or simultaneous double soft graviton theorems because at the leading level this choice is irrelevant in case for gravitons. This choice becomes subtle in QCD.} in the un-dressed states as
\begin{align}\label{ge}\nonumber
\Big(\braout Q_{\mathrm{soft}}(\hat{p}_{1})Q_{\mathrm{soft}}(\hat{p}_{2})&\odress\sa\idress\ketin\Big)_{\mathrm{external}}\\
&=\frac{\kappa^{2}}{4} S^{(0)}(\hat{p}_{1};\{k_{i}\})S^{(0)}(\hat{p}_{2};\{k_{j}\})\braout \odress\sa\idress\ketin .
\end{align}
The second contribution \eqref{do2} can be evaluated in the following way. The soft graviton connected to the external particle can be evaluated by single leading soft graviton theorem while the soft graviton connected to the dressing operator can be evaluated by the contraction of the soft operator with the dressing, i.e.,  $[Q_{\mathrm{soft}}(\hat{p}),e^{R_{N}}]$. Therefore this contribution can be written as
\begin{figure}
	\includegraphics[width=\linewidth]{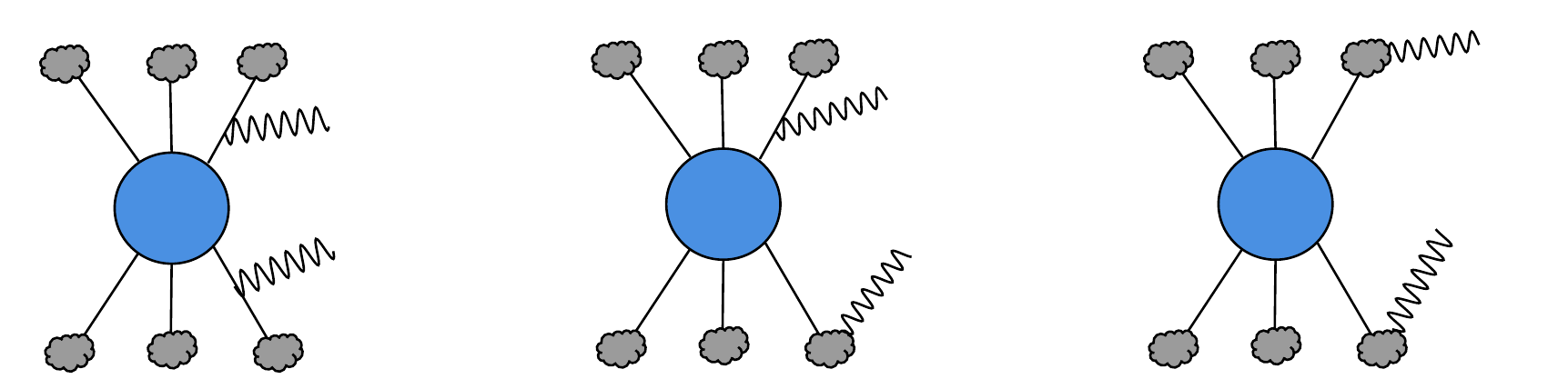}
	\caption{Diagrams illustrate the different ways to connect two soft graviton modes in the Feynman diagram. The first diagram shows both soft gravitons connected to the external legs. The second diagram shows one of the soft graviton connected to the external leg while the other connected to the dressing operator. The last diagram involves both soft gravitons connected to the dressing operator.}
	\label{fig:boat1}
\end{figure}

\begin{align}\nonumber
\frac{\kappa}{2}&S^{(0)}(\hat{p}_{1};{k_{i}})\Big(\braout[Q_{\mathrm{soft}}(\hat{p}_{2}),\odress]\sa\idress\ketin + \braout \odress \sa[Q_{\mathrm{soft}}(\hat{p}_{2}),\idress]\ketin\Big)+\\
&~~~\frac{\kappa}{2}S^{(0)}(\hat{p}_{2};\{k_{i}\})\Big(\braout[Q_{\mathrm{soft}}(\hat{p}_{1}),\odress]\sa\idress\ketin + \braout \odress\sa[Q_{\mathrm{soft}}(\hat{p}_{1}),\idress]\ketin\Big).
\end{align}
Using the form of the dressing operator we derived in the previous section we can finally evaluate the above expression to 
\begin{align}\label{gec}\nonumber
\Big(\braout Q_{\mathrm{soft}}(\hat{p}_{1}) Q_{\mathrm{soft}}(\hat{p}_{2})\odress&\sa\idress\ketin\Big)_{\mathrm{external}+\mathrm{dressing}}=\\
&-\frac{\kappa^{2}}{2}S^{(0)}(\hat{p}_{1};\{k_{i}\})S^{(0)}(\hat{p}_{2};\{k_{j}\})\braout \odress\sa\idress\ketin .
\end{align}
\\
The final contribution \eqref{do3} can be evaluated by the contraction of the soft operators with the dressing. Hence this term can be written as
\begin{align}\nonumber
&\braout[Q_{\mathrm{soft}}(\hat{p}_{1}),[Q_{\mathrm{soft}}(\hat{p}_{2}),\odress]]\sa\idress\ketin +\braout[Q_{\mathrm{soft}}(\hat{p}_{1}),\odress]\sa[Q_{\mathrm{soft}}(\hat{p}_{2}),\idress]\ketin\\
&~~+\braout[Q_{\mathrm{soft}}(\hat{p}_{2}),\odress]\sa[Q_{\mathrm{soft}}(\hat{p}_{1}),\idress]\ketin+\braout\odress\sa[Q_{\mathrm{soft}}(\hat{p}_{1}),[Q_{\mathrm{soft}}(\hat{p}_{2}),\idress]]\ketin.
\end{align}
which evaluates to
\begin{align}\label{gc}\nonumber
\Big(\braout Q_{\mathrm{soft}}(\hat{p}_{1})Q_{\mathrm{soft}}(\hat{p}_{2})\odress&\sa\idress\ketin\Big)_{\mathrm{dressing}}=\\
&~~~~~~\frac{\kappa^{2}}{4} S^{(0)}(\hat{p}_{1};\{k_{i}\})S^{(0)}(\hat{p}_{2};\{k_{j}\})\braout \odress\sa\idress\ketin .
\end{align}
\\
Adding up all the contributions \eqref{ge}, \eqref{gec} and \eqref{gc} we finally get
\begin{align}
\braout Q_{\mathrm{soft}}(\hat{p}_{1})Q_{\mathrm{soft}}(\hat{p}_{2})\odress\sa\idress\ketin=0.
\end{align}

Therefore as expected the two soft gravitons modes indeed decouple, if we are using the dressing operator \eqref{lgdress}.  Without loss of generality one can extend this analysis to multiple soft graviton modes and arrive at the same result, which suggests that the assumption \eqref{st2} holds as expected. 

As an aside we can also use the orthogonality for one soft graviton mode and single soft graviton theorem to constrain the dressing ansatz we had started with. This is demonstrated in appendix \ref{gra1}.
In short, in perturbative gravity it turns out that by the super translation Ward identity and the condition of orthogonality in the dressed states, the form of dressed state is fixed. Additionally one can use the orthogonality conditon in the dressed states and the single soft theorem in the un-dressed states to constrain the dressing (appendix \ref{gra1}). Both of these constraints lead to the same conclusion.
\subsection{An Alternative Dressing}\label{gr6}
%\subsection{2d dressing}
In all the previous  analysis, for the determination of dressed states, we assumed \eqref{st41} in which the soft gravitons are treated as zero frequency limit of finite energy gravitons. This together with other assumptions \eqref{st1}, \eqref{st2} and \eqref{st3}, one naturally obtains the Faddeev-Kulish states. But as pointed out by the authors in \cite{strom}, for obtaining the correct BMS transformations, one needs to treat the soft modes as independant degrees of freedom. This is equivalent to extending the radiative phase space, so that one includes not only the free data $C_{zz}$ and $C_{\bar{z}\bar{z}}$\footnote{$C_{zz}$, $C_{\bar{z}\bar{z}}$ are the radiative metric components of an asymptotically flat metric. These are unconstrained by Einstein equations and all other radiative components can be written in terms of $C_{zz}$ and $C_{\bar{z}\bar{z}}$ .}, but also the boundary modes (boundary of null infinity)  defined by $C(z,\bar{z})$ and $N(z,\bar{z})$. These additional data are defined by
\begin{align}
C_{zz}\mid_{\mathcal{I}^{+}_{-}}&=D_{z}^{2}C,\\
\int_{-\infty}^{\infty}du N_{zz}&=D_{z}^{2}N.
\end{align}
where $N_{zz}=\partial_{u}C_{zz}$ is Bondi news tensor. $D_{z}$ refers to the covariant derivative w.r.t to $2$-sphere. After taking this into account the action of BMS supertranslation charge ($Q[f]$) on these data can be written as
\begin{align}\nonumber\label{db}
[Q[f],N_{zz}(u,z,\bar{z})]&=f(z,\bar{z})~\partial_{u}N_{zz}(u,z,\bar{z}),\\\nonumber
[Q[f],C_{zz}(u,z,\bar{z}]&=f(z,\bar{z})~\partial_{u}C_{zz}(u,z,\bar{z})-2 D_{z}^{2}f(z,\bar{z}),\\\nonumber
[Q[f],N(z,\bar{z})]&=0,\\
[Q[f],C(z,\bar{z})]&=-2f(z,\bar{z}).
\end{align}
\\
In light of the above, we would additionaly like to use the assumptions \eqref{st1}, \eqref{st2} and \eqref{st3} to determine a set of dressed states. 
Let us consider an ansatz for the dressed state of the following form
\begin{align}
e^{R_{N}}=\exp\Big(\frac{\kappa}{2}\int d[k]~\rho(k)\int d^{2}\hat{q}~ D_{\bar{z}}^{2}N(\hat{q};k)  C(z,\bar{z})\Big).
\end{align}
where $\rho (k)=b^{\dagger}(k)b(k)$ is the number operator for the external particles\footnote{we consider massless scalar particles in this case for convenience}. Here $(z,\bar{z})$ are the coordinates for representing the direction $q$ in the integral over the sphere. $N(\hat{q};k)$ is an arbitrary real function and $D_{\bar{z}}$ refers to the covariant derivative w.r.t 2-sphere.

The action of the dressing operator on a ``bare" state is similar to \eqref{dress} and we can write the dressed state as
\begin{align}\label{dress2}
e^{R_{N}}\ket{k}=e^{R_{N}}b^{\dagger}(k)\ket{0}=[e^{R_{N}},b^{\dagger}(k)]\ket{0}= e^{R_{N}(k)}b^{\dagger}(k)\ket{0}.
\end{align}
where
\begin{align}\label{gansatz2}
e^{R_{N}(k)}=\exp{\Big(\frac{\kappa}{2} \int d^{2}\hat{q}~D_{\bar{z}}^{2} N(\hat{q};k_{i})  C(z,\bar{z})\Big)}.
\end{align}
\\
In this section, for clarity let us start with the supertranslation charge $Q[f]$ for an arbitrary $f(z,\bar{z})$ and then later on in the calculations we will substitute for the particular $f(z,\bar{z})$ in \eqref{leadfn}.
The supertranslation Ward identity between such states can be written as
\begin{align}
\braout\odress[Q[f],\sa]\idress\ketin=0.
\end{align}
where
\begin{align}
\odress&=\exp{\Big(-\frac{\kappa}{2} \sum_{i =\mathrm{out}}\int d^{2}\hat{q}~ D_{\bar{z}}^{2}N_{i}(\hat{q};k_{i})C(z,\bar{z})\Big)},\\
\idress&=\exp{\Big(\frac{\kappa}{2} \sum_{i =\mathrm{in}}\int d^{2}\hat{q}~ D_{\bar{z}}^{2}N_{i}(\hat{q};k_{i})C(z,\bar{z})\Big)}.
\end{align}
\\
%\begin{align}
%\braout\odress[Q_{\mathrm{soft}}(\hat{p})+Q_{\mathrm{hard}}(\hat{p}),\sa]\idress\ketin=0
%\end{align}
%where $Q(\hat{p})=Q_{\mathrm{soft}}(\hat{p})$ and $Q_{\mathrm{hard}}(\hat{p})$ are defined as per \eqref{gscharge2} and \eqref{ghcharge2} respectively.
The above equation can be expanded as
\begin{align}\nonumber
\braout Q[f]&\odress \sa\idress\ketin-\braout\odress\sa\idress Q[f] \ketin+\\
&\braout [\odress,Q[f]]~\sa~\idress\ketin-\braout\odress\sa[Q[f],\idress]\ketin=0.
\end{align}

If we use assumption \eqref{st2} then the hard part of the charge only contributes to first two terms in the above equation. The last two terms can be evaluated using \eqref{db}. After a bit of calculation one arrives at 
\begin{align}\nonumber
\kappa\Big(\sum_{i=\mathrm{out}}\int d^{2}\hat{q}~ N_{i}(\hat{q};k_{i})D_{\bar{z}}^{2}f_{i}(\hat{q})&-\sum_{i=\mathrm{in}}\int d^{2}\hat{q}~ N_{i}(\hat{q};k_{i})D_{\bar{z}}^{2}f_{i}(\hat{q})\Big)\braout\odress\sa\idress\ketin\\
&-\kappa\Big(\sum_{i=\mathrm{out}}f_{i}(\hat{k}_{i})E_{i}-\sum_{i=\mathrm{in}}f_{i}(\hat{k}_{i})E_{i}\Big)\braout\odress\sa\idress\ketin=0.
\end{align}
Now if one chooses $f(z,\bar{z})$ in \eqref{leadfn}, so that the integral over $\hat{q}$ gets localised in a particular direction say $\hat{p}$, then the above equation simplifies to 
\begin{align}\nonumber
\kappa\Big(\sum_{i=\mathrm{out}}N_{i}(\hat{p};k_{i})-&\sum_{i=\mathrm{in}}N_{i}(\hat{p};k_{i})\Big)\braout\odress\sa\idress\ketin\\
&-\frac{\kappa}{2}\Big(\sum_{i=\mathrm{out}}S^{(0)}(\hat{p};k_{i})-\sum_{i=\mathrm{in}}S^{(0)}(\hat{p};k_{i})\Big)\braout\odress\sa\idress\ketin=0.
\end{align}
\\
We can now use the assumption \eqref{st3} to get the constraint
\begin{align}
\frac{\kappa}{2}\Big(\sum_{i=\mathrm{out}}S^{(0)}(\hat{p};k_{i})-\sum_{i=\mathrm{in}}S^{(0)}(\hat{p};k_{i})\Big)=\kappa\Big(\sum_{i=\mathrm{out}}N_{i}(\hat{p};k_{i})-\sum_{i=\mathrm{in}}N_{i}(\hat{p};k_{i})\Big).
\end{align}
\\
From the above equation it is natural to associate
\begin{align}
N_{i}(\hat{p};k_{i})=\frac{1}{2}S^{(0)}(\hat{p};k_{i})=\frac{\epsilon^{+}_{\mu\nu}(\hat{p})~k_{i}^{\mu}k_{i}^{\nu}}{2 (p/E_{p}).k_{i}}.
\end{align}
Here $\epsilon^{+}_{\mu\nu}(\hat{p})=\epsilon^{+}_{\mu}(\hat{p})\epsilon^{+}_{\nu}(\hat{p})$.
Therefore the dressed state can be written as
\begin{align}
e^{R_{N}}\ket{k}=e^{R_{N}(k)}\ket{k}=\exp\Big(\frac{\kappa}{2}\int d^{2}\hat{q}~D_{\bar{z}}^{2}~\Big(\frac{\epsilon^{+}_{\mu\nu}(\hat{q})~k^{\mu}k^{\nu}}{ (q/E_{q}).k}~\Big)~C(z,\bar{z})\Big)\ket{k}.
\end{align}
\\
We can again simplify the above expression due the property
\begin{align}
D_{\bar{z}}^{2}~\Big(\frac{\epsilon^{+}_{\mu\nu}(\hat{q})~k^{\mu}k^{\nu}}{ (q/E_{q}).k}~\Big)=(2\pi)E_{k}\delta^{(2)}(z-z_{k}).
\end{align}
\\
where $( z_{k},\bar{z}_{k} )$ refers to direction of the external particle. Finally the dressed particle can be written as
\begin{align}
e^{R_{N}(k)}\ket{k}=\exp\Big(\frac{\kappa}{2}~E_{k}~C(z_{k},\bar{z}_{k})\Big)\ket{k}.
\end{align} 
It is important to note that unlike the Faddeev-Kulish states, these dressed states are made out of Goldstone modes conjugate to zero modes. Unlike the derivation for Faddeev-Kulish states from supertranslation Ward identity in which one has to neglect the $\mathcal{O}(\Lambda)$ terms where $\Lambda$ is the upper cut off for the dressing, one need not make this assumption in this present derivation. The relationship between such states and the Faddeev-Kulish states has been analysed in \cite{strom-fadeev}.

\section{From Asymptotic Symmetries to Dressed states in QCD}\label{qcd}

In this section we apply the same analysis as we have done to the case of gravity. We try to find a set of dressed states that are compatible with the Ward identities associated to large gauge transformations in Yang-Mills theory. In section \ref{qcd31}, the equivalence of Ward identities of large gauge transformations with leading soft gluon theorem is reviewed. In section \ref{qcd32} we start with a simple ansatz (inspired from gravity) for a dressed state in QCD and  constrain the dressing using certain assumptions which we will discuss later in the section. As a check we also verify the orthogonality condition for the dressing operator using the soft gluon theorem in section \ref{qcd33}.

\subsection{Review of Single Soft Gluon Theorem in Undressed States} \label{qcd31}

Here we briefly review the equivalence between leading single soft gluon theorem and asymptotic symmetries in Yang-Mills theory. The reader can refer to \cite{strom-ym1,strom-ym2,Note_Mao} for further details. The asymptotic charge defined at future null infinity can be written as 
\begin{align}\label{ymcharge1}
Q[\alpha]=\frac{1}{g^{2}}\int_{\mathcal{I}^{+}_{-}}\mathrm{tr}(\alpha* \mathcal{F}).
\end{align}
where $ \mathcal{I}^{+}_{-}$ is the past of future null infinity.
$\mathcal{F}$ is the gauge field strength defined as $\mathcal{F}=\mathcal{F}_{\mu\nu}^{a}T^{a}$, where $\mathcal{F}_{\mu\nu}^{a}=\partial_{\mu}\mathcal{A}_{\nu}^{a}-\partial_{\nu}\mathcal{A}_{\mu}^{a}+g f^{abc}\mathcal{A}_{\mu}^{b}\mathcal{A}_{\nu}^{c}$. $\alpha\equiv\alpha^{a}(z,\bar{z})T^{a}$, where  $\alpha^{a}(z,\bar{z}) $ is an arbitrary function on the sphere and $T^{a}$ refers to the Lie algebra generator. $g$ is the coupling constant. Using equation of motion ($D^{\mu}F_{\mu\nu}=g^{2} j_{\nu}$, where $D^{\mu}$ is the gauge covariant derivative and $j_{\nu}$ the matter current) and the asymptotic fall offs of the gauge fields $\mathcal{A}_{\mu}^{a}$'s near null infinity\cite{strom-ym1,strom-ym2,Note_Mao}, we can write the asymptotic charge as  sum of a soft part $	Q_{\mathrm{soft}}[{\alpha}]$ and a hard part 	$Q_{\mathrm{hard}}[{\alpha}]$. After quantisation one can write the soft charge as \cite{strom-ym1}
\begin{align}\label{ymsoft}
Q_{\mathrm{soft}}[{\alpha}]=-\lim\limits_{E_{p} \to{ 0}} E_{p}\int d^{2}w\frac{\sqrt{2}\partial_{\bar{w}}\alpha^{b}(w,\bar{w})}{2(1+w \bar{w})}\Big( a_{+}^{b}(E_{p} ~\hat{p})+~ a_{-}^{b\dagger}(E_{p}~ \hat{p}) \Big),
\\
=-\lim\limits_{E_{p} \to{ 0}} E_{p}\int d^{2}w\frac{\sqrt{2}\partial_{w}\alpha^{b}(w,\bar{w})}{2(1+w \bar{w})}\Big( a_{-}^{b}(E_{p} ~\hat{p})+ a_{+}^{b\dagger}(E_{p} ~\hat{p}) \Big).
\end{align}
Here, $\hat{p}$ refers to the direction of the soft gluon labelled by the coordinates ($w,\bar{w}$).
Using crossing symmetry one can rewrite the soft charge so as to include only one of the polarisations in each of the expressions. Then the above expression can be further written as
\begin{align}\label{ymsoftc}
Q_{\mathrm{soft}}[{\alpha}]=-\lim\limits_{E_{p} \to{ 0}}E_{p}\int d^{2}w\frac{\sqrt{2}\partial_{\bar{w}}\alpha^{b}(w,\bar{w})}{(1+w \bar{w})}a_{+}^{b}(E_{p}~\hat{p}),
\\
=-\lim\limits_{E_{p} \to{ 0}}E_{p} \int d^{2}w\frac{\sqrt{2}\partial_{w}\alpha^{b}(w,\bar{w})}{(1+w \bar{w})}a_{-}^{b}(E_{p}~\hat{p}).
\end{align}
\\
%	Note that,
%	\begin{equation}
%	Q^{\mathrm{soft}}[{\alpha}]\ket{\mathrm{in}}=0 \label{Q_soft in}
%	\end{equation}
Similarly one can write the action of the hard charge on an external state with momentum $k$ and color $c$  as\footnote{The above mentioned action of the hard charge is for tree-level only. In this work we are only considering the action of the hard charge at tree-level and therefore the loop-corrections to the charge are not considered.}
\begin{equation}\label{ymhard}
Q_{\mathrm{hard}}[{\alpha}]\ket{(k,c)}=g \alpha^{b}(\hat{k})~T^{b}_{c d}\ket{(k,d)}.
\end{equation}
where $\hat{k}$ refers to the direction of momentum of the external state.\\
The Ward identity for large gauge transformation can be written as
\begin{align}\label{ymward}
\braout[Q[\alpha],\sa]\ketin=0 \Leftrightarrow 	\braout[Q_{\mathrm{soft}}[\alpha],\sa]\ketin=-\braout[Q_{\mathrm{hard}}[\alpha],\sa]\ketin.
\end{align}
Using the definition of soft and hard charge the above Ward identity becomes,
\begin{align}\nonumber 
\lim\limits_{E_{p} \to{ 0}}E_{p}\int d^{2}w\frac{\sqrt{2}\partial_{w}\alpha^{b}(w,\bar{w})}{(1+w \bar{w})} &\braout a_{+}^{b}(E_{p}~\hat{p}) \sa\ketin\\&=-g\Big[\sum_{\mathrm{out}}\alpha^{b}(\hat{k_{i}})~T_{i}^{b}-\sum_{\mathrm{in}}\alpha^{b}(\hat{k_{i}})~T_{i}^{b} \Big] \braout\sa\ketin.\label{WI}
\end{align}
The leading soft gluon theorem\footnote{We are considering tree-level S-matrix.} \cite{strom-ym1} for a positive helicity soft gluon of color $a$ and in the direction $\hat{p}$ labelled by the coordinates ($w_{p},\bar{w_{p}}$) can be written as
\begin{align}\label{lsgt}
\lim\limits_{E_{p} \to{ 0}}E_{p}\braout a_{+}^{a}(E_{p}~\hat{p})\sa\ketin=
gS^{(0)a}(\hat{p};\{k_i\})\bra{\mathrm{out}}S\ket{\mathrm{in}}.
\end{align}
where,
\begin{equation}
S^{(0)a}(\hat{p};\{k_i\})=\sum_{i =\mathrm{out}}S^{(0)a}(\hat{p};k_i)- \sum_{i =\mathrm{in}}S^{(0)a}(\hat{p};k_i).
\end{equation}
with
\begin{align}\label{gsfactor}
S^{(0)a}(\hat{p};k_i)\equiv\frac{\epsilon^{+}(\hat{{p}})\cdot k_{i}}{(p/E_{p})\cdot k_{i}}T^a_i. 
\end{align}
where $p\equiv E_{p}(1,\hat{p})$ denotes the $4$-momentum of the soft gluon and $\epsilon^{+}(\hat{p})$ refers to the polarisation vector of the soft gluon which is given by $\epsilon^{+}(\hat{{p}})=1/\sqrt{2}(\bar{w_{p}},1,-i,-\bar{w_{p}})$. 
 $k_{i}$ denotes the $4$-momentum and $T^{a}_{i}$ denotes the  Lie algebra generator in the representation of the $i$th hard particle.%
%  If we express the soft factor \eqref{gsfactor} in the sphere coordinates we get
%\begin{align}\label{gsfactor2}
%S^{(0)a}(\hat{p};k_i)=\frac{(1+w_{p}\bar{w_{p}})}{(w_{i}-w_{p})}T^a_i. 
%\end{align}
\\
If one chooses a particular $\alpha(w,\bar{w})$ which is,
\begin{align}
\alpha=\alpha^{b}(w,\bar{w})T^{b}=\frac{\delta^{ab}(1+w_p\bar{w}_p)}{w-w_{p}}T^{b}.\label{gauge parameter}
\end{align}
The Ward identity \eqref{WI} matches with the  leading soft gluon theorem \eqref{lsgt}. For this  particular choice  of $ \alpha$, we denote the soft and hard charges as $Q_{\mathrm{soft}}^a(\hat{p})$ and $Q_{\mathrm{hard}}^{a}(\hat{p})$ respectively and are given by:
\begin{align}\label{gs1}
Q_{\mathrm{soft}}^a(\hat{p})&=\lim\limits_{E_{p} \to{ 0}}E_{p}~ a_{+}^{a}(E_{p}~\hat{{p}}).\\
\label{gh1}
Q_{\mathrm{hard}}^{a}(\hat{p})\ket{(k,b)}&=-g(S^{(0)a}(\hat{p};k) ~)_{b c}\ket{(k,c)}.
\end{align}
\\
We will be using $ Q_{\mathrm{soft}}^a(\hat{p})$ and $ Q_{\mathrm{hard}}^{a}(\hat{p})$ in the rest of our calculations.\\
It is important to note that in this work we are working with tree-level asymptotic charges only. Although we restrict our attention to the case where quarks are massless (as the understanding of QCD asymptotic symmetries is most developed in this context), the analysis of section \ref{qcd4} will not rely on this assumption.\footnote{As an aside, we note that assuming that quarks are massless is not an unreasonable assumption in the context of perturbative QCD where S-matrix is well defined.} We will be working with a gauge group  $SU(N)$ in which the gauge generators satisfy,
\begin{align}
[T^{a},T^{b}]=i f^{abc} T^{c}.
\end{align}
are normalised as $\mathrm{Tr}(T^{a}T^{b})=\frac{1}{2}\delta^{ab}$. The gluons transform in the adjoint representation ($(t^{a})_{bc}=if^{abc}t^{c}$) while quarks transform in the fundamental  representation. 

\subsection{Dressed States from Ward Identity }\label{qcd32}
%There is also another way to obtain the constraints on the dressing operator.
%Our starting point is the same working hypothesis using which we derive the structure of Asymptotic states in for Gravity S matrix.
In section \ref{gr3}, it was shown that under certain assumptions one can recover a set of dressed states in perturbative gravity from asymptotic symmetries. In this section we will proceed along the same lines. We start with the following assumptions:

%We now try to understand the infra-red structure of Dressed states in perturbative
%QCD. Our starting point is the same working hypothesis using which we derive the structure of Dressed states in for quantum gravity S matrix.
%	 In the frame work of  gravity,  we have shown that the Ward Identity along with orthogonality relation leads to Faddeev-Kulish type of dressing. And it was further shown that a finite number of soft graviton modes satisfy Orthogonality condition. Another important assumption was  the energy cut-off in the definition of hard charge. 
%	We start with the following assumptions,
\begin{enumerate}
	\item (Non-Abelian) Large gauge transformation is a symmetry of the S-matrix. Due to the non-Abelian nature of the asymptotic symmetry, this assumption is more subtle than the corresponding assumption in gravity where supertranslations generate an Abelian group.\label{lgt1}
	
We quantify this assumption as a hierarchy of Ward identities 

\begin{equation}
[Q^{a_{1}}(\hat{p}_{1}),\ [Q^{a_{2}}(\hat{p}_{2}),[\dots\ [Q^{a_{n}}(\hat{p}_{n}),\ S]\dots]\ =\ 0\ \forall\ n.
\end{equation}		

where $Q^{a_{1}}(\hat{p}_{1})=Q^{a_{1}}_{\mathrm{soft}}(\hat{p}_{1})+Q^{a_{1}}_{\mathrm{hard}}(\hat{p}_{1})$ are defined in \eqref{gs1} and \eqref{gh1}.
\item Soft gluon modes satisfy orthogonality relations  with respect to asymptotic states of the  theory. In this context orthogonality condition means that finite number of insertion of soft gluons generates a null state (i.e. the resulting state is orthogonal to all the asymptotic states of the theory without soft external gluons). This is equivalent to the equation,
\begin{align}
\braout~ Q^{a_{1}}_{\mathrm{soft}}(\hat{p}_{1})~Q^{a_{2}}_{\mathrm{soft}}(\hat{p}_{2})\ldots Q^{a_{n}}_{\mathrm{soft}}(\hat{p}_{n})~\qout~\sa~\qin~\ketin=0.
\end{align}

where $\qout,\qin$ are the dressing operator acting on the out and in states respectively.

\label{lgt2}
	%		Finite number of soft gluon modes decouple from the S matrix.
%\item The hard charge, $Q_{\mathrm{hard}}^{a}(\hat{p})$, has a trivial action on the dressing and therefore the action on a dressed state coincides with the  action on the undressed state. \label{lgt3}
	%\item  Leading soft gluon theorems in the undressed states for $N \geqslant 2$, where $N$ is the number of gluons.\label{lgt4}
	\item The S-matrix elements are non-trivial.\label{lgt5}
\end{enumerate}

%We note that assumption \ref{lgt1} is more intricate then the corresponding assumption in gravity. More in detail, when we recast it as  Ward identities, it amounts to 
%\begin{equation}
%[Q[\epsilon_{1}],\ [Q[\epsilon_{2}],\ [\dots, [Q_{\epsilon_{n},S]\dots]\ =\ 0
%\end{equation}

%Assumption \ref{lgt4} is a rather crucial assumption in our analysis. As we will argue in section (CITE SECTION)  if our asymptotic states are color neutral then this assumption is indeed satisfied. A more detailed analysis of this assumption remains outside the scope of this paper.  

As a warm up, let us first consider the simplest ansatz for the dressing operator which is motivated from the (infrared finite) asymptotic states of QED and gravity. 

\begin{align}\label{ansatz_QCD}
U_{E}=P_{\bar{E}}\exp \Big(g\int d[k]\rho^{bc}(k)\int^{E}d[q](N^{a\mu}(q;k))_{bc}A_{\mu}^{a}(q)\Big).
\end{align} 
where
\begin{align}\nonumber\label{notations}
&d[k]\equiv \frac{d^{3}k}{(2\pi)^{3}2E_{k}}~~~ ,~~~~~~~ d[q]\equiv \frac{d^{3}q}{(2\pi)^{3}2E_{q}},	\\&\rho^{bc}(k)\equiv b^{\dagger b}(k) b^c(k)~~~,~~~~~ A_{\mu}^{a}(q)\equiv a_{\mu}^{a \dagger}(q)-a_{\mu}^{a}(q).			
\end{align}
Here $k$ refers to the external particle momenta while  $q$ refers to the gluon momenta. 
$b^{\dagger b}(k)$($b^b(k) $)  is the creation (annihilation) operator for the external particle with momentum $ k$ and color $b$. $a_{\mu}^{a \dagger}(q)$, $a_{\mu}^{a}(q) $  are the creation and annihilation operators associated to the gluon field which can be written in the polarisation basis as,
\begin{equation}
a_{\mu}^{a \dagger}(q)=\sum_{r=\pm} \epsilon_{\mu}^r a^{a \dagger}_r(q)\quad \mathrm{and} \quad a_{\mu}^{a}(q)=\sum_{r=\pm} \epsilon_{\mu}^{r*} a^{a}_r(q).
\end{equation}
$a_{r}^a(q)$, $a_{s}^{b \dagger}(q')$ satisfy the normalisation condition:
\begin{equation}
[a_{r}^a(q),a_{s}^{b \dagger}(q')]= (2E_q)(2 \pi)^3 \delta_{rs} \delta(\vec{q}-\vec{q}~')\delta^{ab}.
\end{equation}
$\mathcal{N}^{a \mu}(k,q) $ is an arbitrary matrix valued function in the color space of external particles which has a pole in $E_{q}$. There is an upper cut-off $E$ for the dressing operator which ensures that only low energy gluons are included in the dressing. Unlike gravity, due to the non-abelian nature we choose a particular ordering of the operators $A_{\mu}^{a}(q)$ in \eqref{ansatz_QCD}. $A_{\mu}^{a}(q)$'s are ordered in such a way that the lowest energy operator will act first on the external particles. $ \bar{P}_{E}$ denotes this energy ordering.
\\
The action of the dressing operator \eqref{ansatz_QCD} on an external  single particle state with momentum $ k$ and color index $ b$ can be written as,
\begin{equation}
U_E\ket{(k,b)}=(U_{E}(k))_{bc}\ket{(k,c)}.
\end{equation} 
where, $ (U_{E}(k))_{bc}$ is given by
\begin{equation}
(U_{E}(k))_{bc}=\Big(\bar{P}_{E}\exp\Big(g\int^{E}{\mathrm{d}[q]} (~\mathcal{N}^{a \mu}(q;k))~A_{\mu}^{a}(q)\Big)\Big)_{bc}\label{ansats_QCD_multiparticle}.
\end{equation}
In a similar way the action of dressing operator  on multi-particle state can be found to be,
\begin{align}\nonumber
%	\ket{\mathrm{in}}_{\mathrm{d}}=
U_{E}&\ket{{(k_1,b_1),(k_2,b_2)\ldots(k_n,b_n)}}\\&~~~~~~~~~~~~= (U_{E}(k_1))_{b_{1}c_{1}}(U_{E}(k_{2}))_{b_{2}c_{2}}\ldots(U_{E}(k_{n}))_{b_{n}c_{n}}\ket{{(k_1,c_1),(k_2,c_2)\ldots(k_n,c_n)}}.\label{multi dressing}
\end{align}
Therefore the dressing operator ``factorises" in the color space of external particles. In the rest of the calculations we will suppress the color indices and will denote the dressing operator simply by $U_{E}(k_i) $.
%The asymptotic operator  $U_{E}(k)$ given in equation  \eqref{dressed state QCD} is acting on a single particle state $\ket{(k,b)}$. Then according to \eqref{dressed state QCD}, the action of asymptotic operator on a multiparticle state can be given as follows,
%
%\begin{align}
%%	\ket{\mathrm{in}}_{\mathrm{d}}=
%U_{E}\ket{{(k_1,b_1),(k_2,b_2)..,(k_n,b_n)}}=\prod_{k_{i}} U_{E}^{(k_i)}\ket{{(k_1,b_1),(k_2,b_2)..,(k_n,b_n)}}=
%\ket{\mathrm{in}}_{\mathrm{d}}\label{multi dressing}
%\end{align}
%
%where,

%\begin{equation}
%U_{E}=\bar{P}_{E}\exp\Big(g\int^{E}{\mathrm{d}[q_1]}\sum_{k_i=\mathrm{in}} ~\mathcal{N}_{(1)}^{a_1 \mu_1}(k_i,q_1) ~A_{\mu_{1}}^{a_1}(q_1)\Big).\label{dressed QCD multiparticle}

%\end{equation} 
We have defined the dressed state \eqref{ansats_QCD_multiparticle} to have a hierarchy in the softness of gluon momenta in such a way that the $n$th order term in the dressing operator takes the following form:
\begin{equation}\label{eordering}
U_{E}(k)_{{{\rvert}{g^n}}}=~g^n\int^{E}{\mathrm{d}[q_1]}...\int^{E_{q_{n-1}}}{\mathrm{d}[q_n]}\mathcal{N}^{a_1 \mu_1}(q_{1};k)...\mathcal{N}^{a_n \mu_n}(q_n;k) A_{\mu_{1}}^{a_1}(q_1)\ldots A_{\mu_{n}}^{a_n}(q_n).
\end{equation}
The factor ($ n!$) in denominator of expansion of the exponential cancels with the ($ n!$) ways the energy ordering can be taken into account.
\\
Having defined the dressed state let us start with assumption \eqref{lgt1},
\begin{align}
{}_{\mathrm{d}}\bra{\mathrm{out}}[Q^{a}(\hat{p}),\mathrm{S}]\ket{\mathrm{in}}_{\mathrm{d}}=0.
\label{ward identity}
\end{align}
where $Q^{a}(\hat{p})=Q_{\mathrm{soft}}^a(\hat{p})+Q_{\mathrm{hard}}^{a}(\hat{p})$ (which are already defined in  \eqref{gs1} and \eqref{gh1}).\\
$ {}_{\mathrm{d}}\bra{\mathrm{out}}$ and $ \ket{\mathrm{in}}_{\mathrm{d}}$ represents the dressed outgoing and incoming states which  are given by, 
\begin{equation}\nonumber
{}_{\mathrm{d}}\bra{\mathrm{out}}=\bra{\mathrm{out}}U_E^{\dagger\mathrm{out}} \quad \mathrm{and} \quad    \ket{\mathrm{in}}_{\mathrm{d}}=U_E^{\mathrm{in}}\ket{\mathrm{in}}.
\end{equation}
Here,
\begin{equation}\nonumber
U_E^{\dagger\mathrm{out}}=\prod_{k_i=\mathrm{out}}U_E^{\dagger \mathrm{out}}(k_i) \quad \mathrm{and}\quad U_E^{\mathrm{in}}=\prod_{k_i=\mathrm{in}}U_E^\mathrm{in}(k_i).
\end{equation}
\\
After writing the charge as sum of soft and hard part we can write the Ward identity as
\begin{align}
\bra{\mathrm{out}}U_E^{\dagger\mathrm{out}}[Q^{a}_{\mathrm{soft}}(\hat{p}),\mathrm{S}]U_{E}^\mathrm{in}\ket{\mathrm{in}}+\bra{\mathrm{out}}U_E^{\dagger\mathrm{out}}[Q^{a}_{\mathrm{hard}}(\hat{p}),\mathrm{S}]U_{E}^\mathrm{in}\ket{\mathrm{in}}=0.
\label{ward identity1}
\end{align}
The first term can be expanded as,
\begin{align}
\bra{\mathrm{out}}Q_{\mathrm{soft}}^a(\hat{p})U_{E}^{\dagger\mathrm{out}}SU_{E}^{\mathrm{in}}\ket{\mathrm{in}}+\bra{\mathrm{out}}[U_{E}^{\dagger\mathrm{out}},Q_{\mathrm{soft}}^a(\hat{p})]SU_{E}^{\mathrm{in}}\ket{\mathrm{in}}-\bra{\mathrm{out}}U_{E}^{\dagger\mathrm{out}}S[Q_{\mathrm{soft}}^a(\hat{p}),U_{E}^{\mathrm{in}}]\ket{\mathrm{in}}.
\label{ward identity soft1}
\end{align}
Now by assumption \eqref{lgt2} the first term in the above expression vanishes. Hence, \eqref{ward identity soft1} becomes,
\begin{equation}
\bra{\mathrm{out}}[U_{E}^{\dagger\mathrm{out}},Q_{\mathrm{soft}}^a(\hat{p})]SU_{E}^{\mathrm{in}}\ket{\mathrm{in}}-\bra{\mathrm{out}}U_{E}^{\dagger\mathrm{out}}S[Q_{\mathrm{soft}}^a(\hat{p}),U_{E}^{\mathrm{in}}]\ket{\mathrm{in}}.\label{QCD_soft1}
\end{equation}
The commutator involved in the above expression has been evaluated in appendix \ref{qcda1}, resulting in the following expression: 
\begin{align}\label{softmod}
\bra{\mathrm{out}}U_{E}^{\dagger\mathrm{out}}[Q_{\mathrm{soft}}^a(\hat{p}),S]U_{E}^\mathrm{in}\ket{\mathrm{in}}=g\lim\limits_{E_p \to{ 0}}E_p\Big(\mathcal{N}^a_\mathrm{out}(p;\{k_i\})- \mathcal{N}^a_\mathrm{in}(p;\{k_i\})\Big)  \bra{\mathrm{out}}U_{E}^{\dagger\mathrm{out}}SU_{E}^\mathrm{in}\ket{\mathrm{in}}.
\end{align}
where:
\begin{equation}\label{nout}
\mathcal{N}^{a}_\mathrm{out}(p;\{k_i\})=\sum_{i =\mathrm{out}}\mathcal{N}^{a \mu}(p;k_i)~\epsilon_{\mu}^+(\hat{p}) \quad \mathrm{and}  \quad \mathcal{N}^a_\mathrm{in}(p;\{k_i\})=\sum_{i =\mathrm{in}}~\mathcal{N}^{a \mu}(p;k_i)~\epsilon_{\mu}^{+}(\hat{p}).
\end{equation}
%\begin{align}
%\bra{\mathrm{out}}[U_{E}^{\dagger\mathrm{out}},Q_{\mathrm{soft}}^a(\hat{p})]SU_{E}^{\mathrm{in}}\ket{\mathrm{in}}=-g \lim\limits_{E_p \to 0} E_p~\mathcal{N}^{a}_\mathrm{out}(p;\{k_i\}) ~\bra{\mathrm{out}}U_{E}^{\dagger\mathrm{out}}SU_{E}^{\mathrm{in}}\ket{\mathrm{in}}\label{commutator1}
%\end{align}
%and,
%\begin{align}
%\bra{\mathrm{out}}U_{E}^{\dagger\mathrm{out}}S[Q_{\mathrm{soft}}^a(\hat{p}),U_{E}^{\mathrm{in}}]\ket{\mathrm{in}}=-g ~\lim\limits_{E_p \to 0} E_p~\mathcal{N}^a_\mathrm{in}(p;\{k_i\}) \bra{\mathrm{out}}U_{E}^{\dagger\mathrm{out}}SU_{E}^\mathrm{in}\ket{\mathrm{in}}\label{commutator3}
%\end{align}
%where,
%\begin{equation}
%\mathcal{N}^{a}_\mathrm{out}(p;\{k_i\})=\sum_{i =\mathrm{out}}\mathcal{N}^{a \mu}(p;k_i)~\epsilon_{\mu}^+(p) \quad \mathrm{and}  \quad \mathcal{N}^a_\mathrm{in}(p;\{k_i\})=\sum_{i =\mathrm{in}}~\mathcal{N}^{a \mu}(p;k_i)~\epsilon_{\mu}^+(p).
%\end{equation}
%	\begin{equation}
%	\bra{p}^{b}T^{*a}~U_{E}^{\dagger (p)}=\bra{p}^{\beta}{(T^{*a})_{b\alpha}}{({U_{E}}{^{\dagger(p)}})}_{\alpha \beta}.
%	\end{equation}\\
%Then \eqref{QCD_soft1} reduces to, 
%\begin{align}\nonumber
%\bra{\mathrm{out}}U_{E}^{\dagger\mathrm{out}}[Q_{\mathrm{soft}}^a(\hat{p}),S]U_{E}^\mathrm{in}\ket{\mathrm{in}}=-g\lim\limits_{E_p \to{ 0}}E_p\Big(\mathcal{N}^a_\mathrm{out}(p;\{k_i\})- \mathcal{N}^a_\mathrm{in}(p;\{k_i\})\Big)  \bra{\mathrm{out}}U_{E}^{\dagger\mathrm{out}}SU_{E}^\mathrm{in}\ket{\mathrm{in}}
%\end{align}
\\
Now let us consider the second term in equation \eqref{ward identity1}, which can be expanded as,
\begin{align}\nonumber
\bra{\mathrm{out}}Q_{\mathrm{hard}}^{a}(\hat{p})U_{E}^{\dagger\mathrm{out}}&SU_{E}^{\mathrm{in}}\ket{\mathrm{in}}+\bra{\mathrm{out}}[U_{E}^{\dagger\mathrm{out}},Q_{\mathrm{hard}}^{a}(\hat{p})]SU_{E}^{\mathrm{in}}\ket{\mathrm{in}}\\ &-\bra{\mathrm{out}}U_{E}^{\dagger \mathrm{out}}SU_{E}^\mathrm{in}Q_{\mathrm{hard}}^{a}(\hat{p})\ket{\mathrm{in}}-\bra{\mathrm{out}}U_{E}^{\dagger\mathrm{out}}S[Q_{\mathrm{hard}}^{a}(\hat{p}),\ U_{E}^{\mathrm{in}}]\ \ket{\mathrm{in}}.
\label{ward identity hard}
\end{align}
The second and fourth terms involve the action of $Q^{a}_{\textrm{hard}}(\hat{p})$ on the dressing operator. This action is non-trivial as the hard charge acts on gluon of arbitrary energy by rotating its color. However as can be readily verified, the action of $Q_{\textrm{hard}}^{a}(\hat{p})$ on the dressing will produce terms which are one order higher in the coupling $g$ as compared to the rest of the terms. Thus if we are interested in determining the dressing at leading order in $g$, these terms vanish. Without a perturbative expansion in $g$, the action of the hard charge on dressing is non-trivial and we will come back to this issue in section \ref{qcd44}. 

 The remaining terms can be evaluated by the action of the hard charge on the external state given by \eqref{gh1}. Therefore we get
\begin{align}
\bra{\mathrm{out}}U_{E}^{\dagger\mathrm{out}}[Q_{\mathrm{hard}}^{a}(\hat{p}),S]U_{E}^{\mathrm{in}}\ket{\mathrm{in}}=-gS^{(0)a}(\hat{p};\{k_i\})\bra{\mathrm{out}}~U_{E}^{\dagger\mathrm{out}}~SU_{E}^{\mathrm{in}}\ket{\mathrm{in}}.
\label{Ward_Identity}
\end{align}
Using \eqref{softmod} and \eqref{Ward_Identity} the Ward identity finally becomes,
\begin{align}\label{scl}
g\Big[S^{(0)a}(\hat{p};\{k_i\})-\lim\limits_{E_p \to{ 0}}E_p \Big(\mathcal{N}^a_\mathrm{out}(p;\{k_i\})- \mathcal{N}^a_\mathrm{in}(p;\{k_i\})\Big)\Big] \bra{\mathrm{out}}U_{E}^{\dagger\mathrm{out}}SU_{E}^{\mathrm{in}}\ket{\mathrm{in}}=0 .
\end{align}
Now we use the last assumption \eqref{lgt5} and hence,
\begin{align}
\sum_{i =\mathrm{out}}S^{(0)a}(\hat{p};k_i)- \sum_{i =\mathrm{in}}S^{(0)a}(\hat{p};k_i)-\lim\limits_{E_p \to{ 0}}E_p \Big(\mathcal{N}^a_\mathrm{out}(p;\{k_i\})- \mathcal{N}^a_\mathrm{in}(p;\{k_i\})\Big)  =0 .\label{QCDconservation_law}
\end{align}
which can be also written as
%\begin{equation}
%	\sum_{k_i =\mathrm{out}}\frac{p \cdot \epsilon^{+}}{k_i \cdot p}- \sum_{k_i =\mathrm{in}}S^{(0)}(a,\hat{p},k_i)-\lim\limits_{E_p \to{ 0}}E_p\Big(\sum_{k_i =\mathrm{out}}\mathcal{N}^{a \mu*}(k_i,p)+\sum_{k_i =\mathrm{in}}~\mathcal{N}_{(1)}^{a \mu}(k_i,p)\Big)~\epsilon^+_{\mu}(p)=0
%\end{equation} 
\begin{align}\nonumber
\lim\limits_{E_p \to{ 0}}E_p\Big(\sum_{i =\mathrm{out}}\mathcal{N}^{a \mu}(p;k_i)-\sum_{i =\mathrm{in}}~\mathcal{N}^{a \mu}(p;k_i)\Big)~\epsilon^+_{\mu}(\hat{p})=\sum_{i =\mathrm{out}}\frac{k_i \cdot \epsilon^{+}(\hat{p})}{k_i \cdot (p/E_{p})}T^{a}_i-\sum_{i =\mathrm{in}}\frac{k_i \cdot \epsilon^{+}(\hat{p})}{k_i \cdot (p/E_{p})}T^{a}_i.
\end{align}
As in gravity we can now  associate,
\begin{equation}
\lim\limits_{E_p \to{ 0}}E_p~\mathcal{N}^{a \mu}(p;k_i)\epsilon^+_{\mu}(\hat{p})=\frac{k_i \cdot \epsilon^{+}(\hat{p})}{k_i \cdot (p/E_{p})}T^{a}_{i}.
\label{final_Q1}
\end{equation} 
Hence
\begin{equation}
\mathcal{N}^{a \mu}(p;k_i) =\frac{k_{i}^{\mu}}{k_i \cdot p}T^{a}_{i}.
\end{equation}
%Note that $\mathcal{N}_{(1)}^{a \mu}(k_i,p)$ has a pole in soft momentum.
\\
Substituting the above expression in the dressing ansatz we finally get the dressed state as,
\begin{equation}
U_E\ket{(k,b)}=\bar{P}_{E}\exp\Big(g\int^{E}{\mathrm{d}[q] } ~\frac{k^\mu}{k \cdot q} T^{a}_k~A_{\mu}^{a}(q)\Big)\ket{(k,b)}. \label{final dressed state1}
\end{equation}

\subsection{Orthogonality Relations for Single Soft Gluon Insertion}\label{qcd33}

We now ask, if this dressed state \eqref{final dressed state1} is such that a finite number of soft gluon modes satisfy the orthogonality relation \eqref{lgt2}. As we show below, this is not the case
and we have to modify the ansatz.
% place an ansatz to improve on the dressing obtained above such that soft modes always decouple from the S-matrix.
\\  
Let us first consider the orthogonality condition involving one soft gluon mode. i.e 

\begin{equation}\label{orthogonality_single}
\bra{\mathrm{out}}Q_{\mathrm{soft}}^a(\hat{p})U_{E}^{\dagger\mathrm{out}}SU_{E}^{\mathrm{in}}\ket{\mathrm{in}}=0 	.
\end{equation}
Using the definition of soft charge, equation \eqref{orthogonality_single} can be written as
\begin{equation}
\lim\limits_{E_p \to{ 0}}E_p~\bra{\mathrm{out}}a_{+}^{a}(E_p ~\hat{p})U_{E}^{\dagger}\mathrm{out}SU_{E}^{\mathrm{in}} \ket{\mathrm{in}}=0.\label{decoupling}
\end{equation} 
%Let us evaluate the l.h.s of \eqref{decoupling}. We start by considering a scattering amplitude with an extra soft gluon mode
%\begin{align}\nonumber
%~\lim\limits_{E_p \to{ 0}}E_p~\bra{\mathrm{out}}a_{+}^{a}(E_p \hat{x_{p}})U_{E}^{\dagger\mathrm{out}}SU_{E}^{\mathrm{in}}\ket{\mathrm{in}}
%\end{align}
The above expression recieve contributions from two terms as in gravity. One of them corresponds to the case when the soft gluon mode is connected to the external legs and the other corresponds to the case when it is connected to the dressing operator. The amplitude with soft gluon mode connected to the external legs is given by leading single soft gluon theorem and it evaluates to,
\begin{equation}
\Big(\bra{\mathrm{out}}Q_{\mathrm{soft}}^a(\hat{p})U_{E}^{\dagger\mathrm{out}}SU_{E}^{\mathrm{in}}\ket{\mathrm{in}}\Big)_{\mathrm{external}}=g  S^{(0)a}(\hat{p};\{k_i\})\bra{\mathrm{out}}~U_{E}^{\dagger\mathrm{out}}~SU_{E}^{\mathrm{in}}\ket{\mathrm{in}} .\label{soft theorem}
\end{equation}
\begin{figure}
	\includegraphics[width=\linewidth]{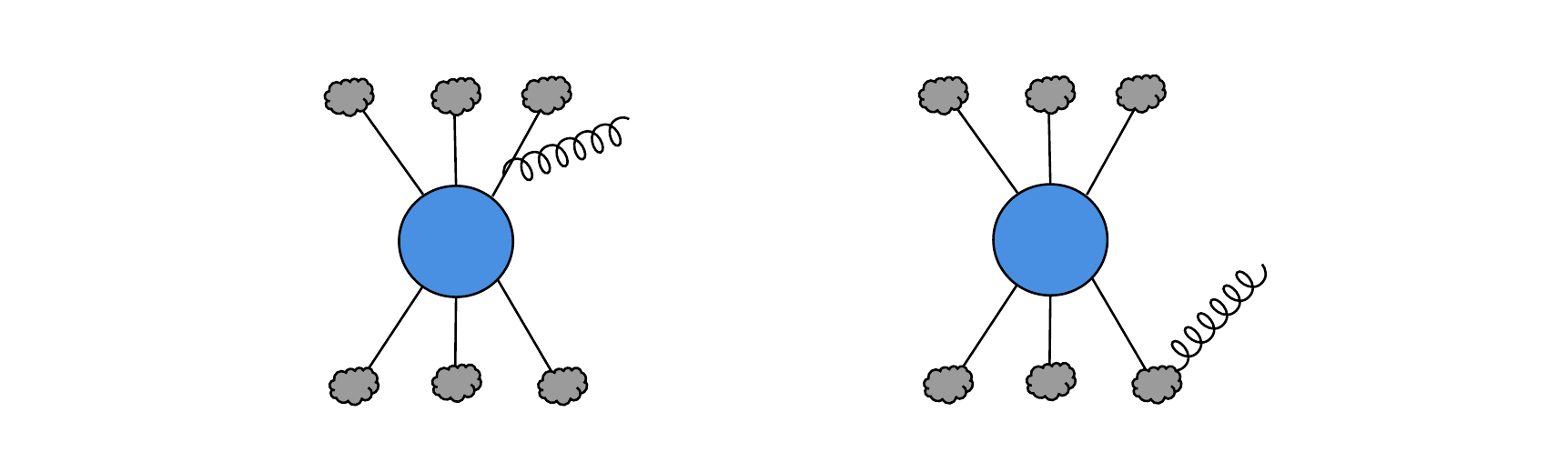}
	\caption{The figure demonstrates the different ways to connect a soft gluon mode in the Feynman diagram. The first diagram involves a soft gluon connected to the external leg while the second involves a soft gluon connected to the dressing operator.}
	\label{fig:gluon}
\end{figure}
When the soft gluon mode is connected to dressing operator the corresponding amplitude can be evaluated by the contraction of the soft operator with the dressing operator,
%		The emmision from the dressing operator can be computed by the constraction of the soft mode with the dressing operator i.e

\begin{align}\nonumber
\Big(\bra{\mathrm{out}}Q_{\mathrm{soft}}^a(\hat{p})&U_{E}^{\dagger\mathrm{out}}SU_{E}^{\mathrm{in}}\ket{\mathrm{in}}\Big)_{\mathrm{dressing}}\\
&=\bra{\mathrm{out}}[Q_{\mathrm{soft}}^a(\hat{p}),U_{E}^{\dagger\mathrm{out}}]~SU_{E}^{\mathrm{in}}\ket{\mathrm{in}}+ \bra{\mathrm{out}}U_{E}^{\dagger\mathrm{out}}~S[Q_{\mathrm{soft}}^a(\hat{p}),U_{E}^{\mathrm{in}}]\ket{\mathrm{in}} .
\end{align}
By using the commutators given in appendix \ref{qcda1} the above term evaluates to,
\begin{align}\label{godress}
\Big(\bra{\mathrm{out}}Q_{\mathrm{soft}}^a(\hat{p})&U_{E}^{\dagger\mathrm{out}}SU_{E}^{\mathrm{in}}\ket{\mathrm{in}}\Big)_{\mathrm{dressing}}=-g  S^{(0)a}(\hat{p};\{k_i\})\bra{\mathrm{out}}~U_{E}^{\dagger\mathrm{out}}~SU_{E}^{\mathrm{in}}\ket{\mathrm{in}}.
\end{align}
%		Using the expression of the dressing operator and the commutator relation we have derived in the appendix this can be computed as
%\begin{align}
%\bra{\mathrm{out}}[U_{E}^{\dagger\mathrm{out}},Q_{\mathrm{soft}}^a(\hat{p})]SU_{E}^{\mathrm{in}}\ket{\mathrm{in}}=- g\sum_{i =\mathrm{out}}~S^{(0)a}(\hat{p};k_i)~\bra{\mathrm{out}}U_{E}^{\dagger\mathrm{out}}SU_{E}^{\mathrm{in}}\ket{\mathrm{in}}.\label{commuatator1}
%\end{align}
\\
Summing \eqref{soft theorem} and \eqref{godress} we finally get
\begin{align}
g\Big[S^{(0)a}(\hat{p};\{k_i\})-S^{(0)a}(\hat{p};\{k_i\})\Big] \bra{\mathrm{out}}U_{E}^{\dagger\mathrm{out}}SU_{E}^{\mathrm{in}}\ket{\mathrm{in}}=0 .\label{single orthogonality final}
\end{align}
\\
We can see that the orthogonality relation involving one soft gluon mode  is satisfied.

However it is rather easy to see that these states do not satisfy orthogonality condition when more then one soft gluon insertions are present. This is simply because such an orthogonality would imply that no consecutive soft emissions of gluons (that is emission of one soft gluon from another soft gluon) can take place. In the next section we start with a rather general ansatz and show that the assumptions \eqref{lgt1}, \eqref{lgt2} and \eqref{lgt5} will lead to asymptotic states of the theory in which soft gluon correlations are taken into account.

\section{Generalised Dressing Operator}\label{qcd4}
As explained in the previous section the dressing ansatz we started with does not include gluon gluon correlations and hence cannot satisfy the orthogonality condition involving multiple soft gluon insertions. In this section we modify the ansatz so as to take this into account.
\\
The most natural modification of the ansatz of the previous section is given by, 
\begin{align}
U_{E}\ket{(k,b)}=(U_{E}(k))_{bc}\ket{(k,c)}.
\end{align}
where
\begin{align}\nonumber
&(U_{E}(k))_{bc}=\Big(\bar{P}_{E}\exp\Big(g \int^{E}{\mathrm{d}[q_1]}  \mathcal{N}_{(1)}^{a_1 \mu_1}(q_1;k) A_{\mu_{1}}^{a_1}(q_1)\\\nonumber&~~~~~~~~~~~~~~~~~~~~~~~~~~~~+g^2 \int^{E}{\mathrm{d}[q_1]}\int^{E_{q_1}}{\mathrm{d}[q_2]}\mathcal{N}_{(2)}^{a_2 \mu_2,a_1 \mu_1}(q_{1},q_{2};k)A_{\mu_{1}}^{a_1}(q_1)A_{\mu_{2}}^{a_2}(q_2)\Big)\\&+g^3 \int^{E}{\mathrm{d}[q_1]}\int^{E_{q_1}}{\mathrm{d}[q_2]}\int^{E_{q_2}}{\mathrm{d}[q_3]}\mathcal{N}_{(3)}^{a_3 \mu_3,a_2 \mu_2,a_1 \mu_1}(q_{1},q_{2},q_{3};k)A_{\mu_{1}}^{a_1}(q_1)A_{\mu_{2}}^{a_2}(q_2)A_{\mu_{3}}^{a_3}(q_3)\Big)+...\Big)_{bc}.\label{dressing QCD modified}
\end{align}
\\
where $\mathcal{N}_{(1)}^{a_1 \mu_1}(q_1;k), \ \mathcal{N}_{(2)}^{a_2 \mu_2,a_1 \mu_1}(q_{1},q_{2};k),\ \mathcal{N}_{(3)}^{a_3 \mu_3,a_2 \mu_2,a_1 \mu_1}(q_{1},q_{2},q_{3};k)\ldots$ are all matrix valued functions in the color space of the external state. The rest of the notations follow from the initial ansatz \eqref{notations}.
\\
The first term in the modified dressed state corresponds to the initial dressing we had started with. In order to include the gluon-gluon correlations we consider an infinite series of terms (series in coupling constant $g$) in the exponent. 
\\
As earlier we would like to determine each term in the exponent (at $\mathcal{O}(g^{n})$) using the assumptions \eqref{lgt1}, \eqref{lgt2} and \eqref{lgt5}. In fact in order to determine the term in the dressing at order $g^{n}$ we only require the following.\footnote{We are assuming a particular ordering in energy associated to the soft charge $E_{p_{1}}>E_{p_{2}}>\dots E_{p_{n}}$.}

%Our goal is to show that each term in the dressing exponent (at $O(g^{n})$ can be determined using the assumptions \ref{lgt1},\ \ref{lgt2}. In fact in order to determine the term in the dressing at order $g^{n}$ we only require the following.

\begin{equation}\label{wconstraint}
\begin{array}{lll}
[Q^{a_{1}}(\hat{p}_{1}),[\ Q^{a_{2}}(\hat{p}_{2}),\ \dots [Q^{a_{n}}(\hat{p}_{n}),\ S]\dots]\ =\ 0\ \textrm{at}\ \mathcal{O}(g^{n}),\\
\langle\textrm{out}\vert\ Q^{a_{1}}_{\textrm{soft}}(\hat{p}_{1})\dots\ Q^{a_{n}}_{\textrm{soft}}(\hat{p}_{1})U_{E}^{\dagger}\ S~U_{E}\ketin\ =\ 0\ \textrm{at}\ O(g^{n}).
\end{array}
\end{equation}
\\
Let us start with the Ward identity involving one asymptotic charge and demand that our states are such that the identity is satisfied at $\mathcal{O}(g)$: 
\begin{align}
{}_{\mathrm{d}}\bra{\mathrm{out}}[Q^{a}(\hat{p}),\mathrm{S}]\ket{\mathrm{in}}_{\mathrm{d}}=0,
\label{qcdwi1}
\end{align} 
where the out and in states are now dressed by the modified dressing operator. We repeat the same procedure we had done with the earlier ansatz by writing the charge as sum of soft and hard part and seperately evaluating these terms. \\
The soft part can be expanded as
\begin{align}\nonumber
&\braout\qout[Q_{\mathrm{soft}}^a(\hat{p}),\sa]\qin\ketin=\\
&\bra{\mathrm{out}}Q_{\mathrm{soft}}^a(\hat{p})U_{E}^{\dagger\mathrm{out}}SU_{E}^{\mathrm{in}}\ket{\mathrm{in}}+\bra{\mathrm{out}}[U_{E}^{\dagger\mathrm{out}},Q_{\mathrm{soft}}^a(\hat{p})]SU_{E}^{\mathrm{in}}\ket{\mathrm{in}}-\bra{\mathrm{out}}U_{E}^{\dagger\mathrm{out}}S[Q_{\mathrm{soft}}^a(\hat{p}),U_{E}^{\mathrm{in}}]\ket{\mathrm{in}}.
\label{ward identity soft}
\end{align}
The commutators involved in the term can be evaluated using the results in appendix \ref{qcda1} (\eqref{mswi1} and \eqref{mswi2}) and the above expression evaluates to 
\begin{align}\nonumber
\bra{\mathrm{out}}Q_{\mathrm{soft}}^a(\hat{p})U_{E}^{\dagger\mathrm{out}}SU_{E}^{\mathrm{in}}\ket{\mathrm{in}}  +g\lim\limits_{E_p \to{ 0}}E_p\Big(\mathcal{N}^{a}_{(1)\mathrm{out}}(p;\{k_i\})- \mathcal{N}^{a}_{(1)\mathrm{in}}(p;\{k_i\})\Big)  \bra{\mathrm{out}}U_{E}^{\dagger\mathrm{out}}SU_{E}^\mathrm{in}\ket{\mathrm{in}}\\
+\mathcal{O}(g^{2}).
\end{align}
where $\mathcal{N}^{a}_{(1)\mathrm{out}}(p;\{k_i\}),\mathcal{N}^{a}_{(1)\mathrm{in}}(p;\{k_i\})$ is defined as 

\begin{equation}\label{nout1}
\mathcal{N}^{a}_{(1)\mathrm{out}}(p;\{k_i\})=\sum_{i =\mathrm{out}}\mathcal{N}^{a \mu}_{(1)}(p;k_i)~\epsilon_{\mu}^+(\hat{p}) \quad \mathrm{and}  \quad \mathcal{N}^{a}_{(1)\mathrm{in}}(p;\{k_i\})=\sum_{i =\mathrm{in}}~\mathcal{N}^{a \mu}_{(1)}(p;k_i)~\epsilon_{\mu}^{+}(\hat{p}).
\end{equation}
The hard part of eqn. (\ref{qcdwi1}) evaluates to 
\begin{align}
\braout\qout[Q_{\mathrm{hard}}^a(\hat{p}),\sa]\qin\ketin\ =\ - gS^{(0)a}(\hat{p};\{k_i\})\bra{\mathrm{out}}~U_{E}^{\dagger\mathrm{out}}~SU_{E}^{\mathrm{in}}\ket{\mathrm{in}}.
\end{align} 
In deriving the above equation, we have used the fact that at $O(g)$ the action of $Q_{\textrm{hard}}^{a}(\hat{p})$ on the dressing operator is zero. 
\\
Therefore the Ward identity finally becomes 
\begin{align}\nonumber
g\Big[-S^{(0)a}(\hat{p};\{k_i\})+\lim\limits_{E_p \to{ 0}}E_p \Big(\mathcal{N}^{a}_{(1)\mathrm{out}}(p;\{k_i\})- \mathcal{N}^{a}_{(1)\mathrm{in}}(p;\{k_i\})\Big)\Big] \bra{\mathrm{out}}U_{E}^{\dagger\mathrm{out}}SU_{E}^{\mathrm{in}}\ket{\mathrm{in}}\\
+\bra{\mathrm{out}}Q_{\mathrm{soft}}^a(\hat{p})U_{E}^{\dagger\mathrm{out}}SU_{E}^{\mathrm{in}}\ket{\mathrm{in}} + \mathcal{O}(g^{2})=0 .
\end{align}
Now if we use our assumption that the dressed states are such that the orthogonality condition for one soft gluon mode holds at $\mathcal{O}(g)$  and the assumption \eqref{lgt5} then  we finally get the conservation law
\begin{align}
S^{(0)a}(\hat{p};\{k_i\})-\lim\limits_{E_p \to{ 0}}E_p \Big(\mathcal{N}^{a}_{(1)\mathrm{out}}(p;\{k_i\})- \mathcal{N}^{a}_{(1)\mathrm{in}}(p;\{k_i\}\Big)=0.
\end{align}
which is same as the initial conservation law \eqref{scl} we got with the simplest ansatz we started with. Hence from this conservation law we can write
\begin{align}
\mathcal{N}_{(1)}^{a \mu}(p;k_i) =\frac{k_{i}^{\mu}}{k_i \cdot p}T^{a}_{i}.
\end{align}

\subsection{Nested Ward Identity}\label{qcd41}

Having determined the first term in the dressing ansatz using the assumptions, we now use the same to extract the second order term in the ansatz.
In order to proceed let us consider the Ward identity involving two asymptotic charges: 

\begin{equation}
[Q^{a}(\hat{p}_{1}),[Q^{b}(\hat{p}_{2}),\sa]]\ =\ 0.
\end{equation}
\\
We now use this Ward identity at $\mathcal{O}(g^{2})$ to determine ${\cal N}_{(2)}^{a_{2}\mu_{2},a_{1}\mu_{1}}(q_{1},q_{2};k)$: 
\begin{align}
\braout\qout[Q^{a}(\hat{p}_{1}),[Q^{b}(\hat{p}_{2}),\sa]]\qin\ketin=0.
\end{align}
 The charges can be written as sum of soft and hard part and the above expression becomes
\begin{align}\label{dwi2}
\braout\qout[Q^{a}_{\mathrm{soft}}(\hat{p}_{1})+Q^{a}_{\mathrm{hard}}(\hat{p}_{1}),[Q^{b}_{\mathrm{soft}}(\hat{p}_{2})+Q^{b}_{\mathrm{hard}}(\hat{p}_{2}),S]]\qin\ketin=0.
\end{align}
 where the soft and hard charges are defined in \eqref{gs1},\eqref{gh1}:
\begin{align}\label{qs1}
Q_{\mathrm{soft}}^a(\hat{p}_{1})&=\lim\limits_{E_{p_{1}} \to{ 0}}E_{p_{1}} a_{+}^{a}(E_{p_{1}}~\hat{{p_{1}}}),\\
Q_{\mathrm{hard}}^{a}(\hat{p}_{1})\ket{(k,b)}&=-g(S^{(0)a}(\hat{p}_{1};k) ~)_{b c}\ket{(k,c)},\\
Q_{\mathrm{soft}}^{b}(\hat{p}_{2})&=\lim\limits_{E_{p_{2}} \to{ 0}}E_{p_{2}} a_{+}^{b}(E_{p_{2}}~\hat{{p_{2}}}),\\
Q_{\mathrm{hard}}^{b}(\hat{p}_{2})\ket{(k,c)}&=-g(S^{(0)b}(\hat{p}_{2};k) ~)_{c d}\ket{(k,d)}.
\end{align}
Using the action of charges the expression \eqref{dwi2} finally evaluates to  (the details are given in appendix \eqref{qcda3}):
\begin{align}\nonumber\label{dwicl}
g^2 ~\Big( \lim\limits_{E_{p_2} \to 0} E_{p_2}\lim\limits_{E_{p_1} \to 0} E_{p_1}\mathcal{N}_{(2)}^{a \mu,b \nu }(p_1,p_2;\{k_i\})\epsilon_{\mu}^{+}(\hat{p}_1)\epsilon_{\nu}^{+}(\hat{p}_2)- S^{(0)b}(\hat{p}_{2};p_{1}))_{ac}~S^{(0)c}(\hat{p_1};\{k_{i}\})\Big)\\
\times \bra{\mathrm{out}}U_{E}^{\dagger\mathrm{out}}SU_{E}^{\mathrm{in}} \ket{\mathrm{in}}+\braout\qsa\qsb\qout\sa\qin\ketin+\mathcal{O}(g^{3})=0.
\end{align}

Now we can use the assumption \eqref{lgt2} and demand that the orthogonality involving two soft gluon modes holds at $\mathcal{O}(g^{2})$ then we get the following conservation law.
\begin{align}\label{cldw}
 \lim\limits_{E_{p_2} \to 0} E_{p_2}\lim\limits_{E_{p_1} \to 0} E_{p_1}\mathcal{N}_{(2)}^{a \mu,b \nu }(p_1,p_2;\{k_i\})\epsilon_{\mu}^{+}(\hat{p}_1)\epsilon_{\nu}^{+}(\hat{p}_2) =S^{(0)b}(\hat{p}_{2};p_{1}))_{ac}~S^{(0)c}(\hat{p_1};\{k_{i}\}).
\end{align}
It is important to note that the $S^{(0)b}(\hat{p}_{2};p_{1}))_{ac}$ represents the soft factor associated to one soft gluon being emitted from another soft gluon. Therefore the matrix involved in the soft factor will be in the adjoint representation. Using these we can finally write
\begin{align}
\mathcal{N}_{(2)}^{a \mu,b \nu }(p_1,p_2;k_i)=\frac{p_{2}^{\mu}}{p_{1}.p_{2}}~\frac{k_{i}^{\nu}}{k_{i}.p_{1}}if^{bac}T^{c}_{k_{i}}.
\end{align}
\\
This procedure can be done recursively to extract the $n$th order term in the exponent of the modified dressed state. For this one needs to evaluate the nested Ward identity involving $n$ asymptotic charges along with the orthogonality condition for $n$ soft gluons. It turns out that $\mathcal{N}_{(n)}^{a_{n} \mu_{n},\ldots,a_1 \mu_1}(q_{1},q_{2},\ldots, q_{n};k)$ term will be picked by the commutators involving $n$ soft charges with the dressing operator i.e., $[Q^{a_{1}}_{\mathrm{soft}}(\hat{p}_{1}),\ [Q^{a_{2}}_{\mathrm{soft}}(\hat{p}_{2}),[\ldots\ [Q^{a_{n}}_{\mathrm{soft}}(\hat{p}_{n}),\ U_{E}]\ldots]$ and along with the commutators of hard operator with the soft operator (\ref{hsc}) one is able to determine $\mathcal{N}_{(n)}^{a_{n} \mu_{n},\ldots ,a_1 \mu_1}(q_{1},q_{2},\ldots, q_{n};k)$. For example $\mathcal{N}_{(3)}^{a_3 \mu_3,a_2 \mu_2,a_1 \mu_1}(q_{1},q_{2},q_{3};k)$ can be found to be,
\begin{align}\nonumber
\mathcal{N}_{(3)}^{a_3 \mu_3,a_2 \mu_2,a_1 \mu_1}(q_{1},q_{2},q_{3};k)=\frac{k^{\mu_{1}}}{q_{1}\cdot k}\cdot&\frac{q_{1}^{\mu_{2}}}{q_{2}\cdot q_{1}}\cdot\frac{q_{1}^{\mu_{3}}}{q_{3}\cdot q_{1}}(if^{a_{1}a_{3}a_{4}})(if^{a_{4}a_{2}a_{5}})T^{a_{5}}_{k}\\
&+~\frac{k^{\mu_{1}}}{q_{1}\cdot k}\cdot\frac{q_{1}^{\mu_{2}}}{q_{2}\cdot q_{1}}\cdot\frac{q_{2}^{\mu_{3}}}{q_{3}\cdot q_{2}}(if^{a_{2}a_{3}a_{4}})(if^{a_{1}a_{4}a_{5}})T^{a_{5}}_{k}.
\end{align}\\

Note that from the Ward identities \ref{wconstraint}, only the pole piece of each of the coefficients $\mathcal{N}_{(n)}^{a_{n} \mu_{n},\ldots,a_1 \mu_1}(q_{1},q_{2},\ldots, q_{n};k)$ of the generalised dressing operator can be determined. Also the orthogonality condition in \ref{wconstraint} can be verified using the similar procedure in section \ref{gr5}. One needs to use the consecutive multi-soft gluon theorems inorder to do so. 

It is interesting to note that the ansatz we have started with has already been identified by Catani et al in \cite{catani1}. They showed that such states naturally emerge from the Faddeev-Kulish approach of asymptotic dynamics and the S-matrix between such states are shown to be infra-red finite at the leading order. The properties of such states have been extensively studied in \cite{catani1,catani2,catani3,catani4,catani5,Marchesini,Mueller}.

\subsection{Some Thoughts on  the Color Rotation of Dressing Operator}\label{qcd44}

Throughout this paper, we faced a thorny issue of evaluating the action of $Q_{\textrm{hard}}^{a}(\hat{p})$ on the dressing operator. For the dressed states in gravity, action of $Q_{\textrm{hard}}^{a}(\hat{p})$ on the dressing exponent did not modify the infrared structure of the dressing and in this sense its effect on the dressed states (as far as evaluating S-matrix elements are concerned) could be ignored. In QCD, the issue is far more subtle as {\bf (a)} $Q_{\textrm{hard}}^{a}(\hat{p})$ on the dressing exponent rotates colors of the constituent soft gluons and {\bf (b)} the ``rotated dressing" obtained by action of hard charge has the same infra-red singularity as the original dressing.  

In our analysis we bypassed this issue by considering Ward identities recursively in  the coupling $g$ due to which action of hard charge on the dressing never appeared in our analysis. However we will now like to speculate that precisely due to points {\bf (a)} and {\bf (b)} mentioned above, hard charge action on the dressing is trivial as far as it's contribution to S-matrix elements is concerned.  

For simplicity, we consider the (finite) action of the hard charge as opposed to its infinitesimal action. Namely let us introduce a (formal) group element $U_{h}(g_{\epsilon})\ =\ \exp(i Q_{h}[\epsilon])$.\footnote{Recall that this group is infinite dimensional and hence our definition is rather formal, but suffices for purposes of this section.} Then,

\begin{equation}
U_{h}(g_{\epsilon})\ U_{E}\ U_{h}(g_{\epsilon}^{-1})\ =\ U^{g_{\epsilon}}_{E}.
\end{equation}
\\
where in the new dressing each soft gluon is rotated in the color space by $g_{\epsilon}$. However due to {\bf (a)} and {\bf (b)} and using the analysis of \cite{fk},  we find it plausible that a dressed state obtained from $U^{g_{\epsilon}}_{E}$
will not be in the asymptotic Hilbert space obtained by using the original dressing operator $U_{E}$. Stated differently,

\begin{equation}
\langle\textrm{out}\vert\ U_{E}^{\dagger}\ \sa\ U^{g_{\epsilon}}_{E}\vert \textrm{in}\rangle\ =\ 0.
\end{equation}
 
This is because the exponential suppression due to virtual infrared soft factors is not cancelled by the dressing as the in-coming and out-going states are dressed by inequivalent (color rotated) dressings.  A detailed analysis of these speculations remain outside the scope of the paper.

\section{Conclusions}\label{conclude}

In this paper, we tried to build upon the work of \cite{strom-fadeev,akhoury2,akhoury3,sever} in the context of perturbative QCD S-matrix. Our goal was to find out the extent to which (leading) asymptotic symmetries of QCD can be used to determine the asymptotic states of the theory. In order to do this, we revisited the analysis in gravity \cite{akhoury3} and extracted out the minimal set of conditions under which super-translation symmetries implied existence of dressed states in gravity. We used these ideas to derive the structure of dressed states in QCD from asymptotic Ward identities. Under certain technical conditons, our analysis led us to the conclusion that the  vacuum structure of QCD which is consistent with asymptotic conservation laws is equivalent to that derived by Catani et al in \cite{catani1}. We would once again like to emphasise that our analysis only relied on certain orthogonality conditions (defined in section \ref{qcd4}) and Ward identities. Nowhere did we make use of multiple soft gluon theorems. 

In addition to the issue of understanding the action of hard charge on the dressing operator, many important questions remain open. As we reviewed for perturbative gravity, if we do not start with an ansatz where the dressing operator involves contribution of low frequency modes (as opposed to soft modes which are precisely the zero frequency Goldstone modes) then Ward identity leads us to an alternative dressing operator where the exponent only contains soft modes. This point was discussed in detail in QED \cite{strom-fadeev,sever} where it was shown how such an alternative dressing is related to the well known Faddeev-Kulish dressing. It will be interesting to start with such an ansatz in QCD where dressing only contains soft (zero frequency) gluons and derive the dressing operator. The relationship of such dressings with the generalised coherent states of \cite{catani1} may shed more light on the relationship between infrared structure of QCD and asymptotic symmetries. In this work we have restricted ourselves to charges at tree level. It would be also interesting to include loop-level corrections to the hard charge and its implications on the dressing operator. We would like to address these issues in a future work. 

 \section*{Acknowledgements}
 
We would like to thank Alok Laddha for posing the problem, numerous discussions, and help in writing the manuscript. We would also like to thank Miguel Campiglia, Prahar Mitra, Ashoke Sen and   Biswajit Sahoo for discussions. We are thankful to  Sujay Ashok for guidance and help with the manuscript. We are thankful to Renjan Rajan John, Sruthy Murali and A.Manu for constant encouragement and support. A.A.H  would like to thank Sudipta Sarkar and IIT Gandhinagar for their hospitality during the completion of this work.

\appendix

\section{Dressed States from Soft Theorem in Gravity}\label{gra1}
In this appendix we will use the orthogonality condition involving one soft graviton mode as a constraint for determining the dressing operator.  In other words we will the study the implications of this orthogonality condition on the ansatz $e^{R_{N}}$.  
\\
We start with the orthogonality condition  
\begin{align}\label{sorthogonal}
\braout Q_{\mathrm{soft}}(\hat{p})\odress\sa\idress\ketin=0.
\end{align}
where $e^{R_{N}}$ is defined as \eqref{dresso}.

Note that l.h.s of this expression corresponds to two terms. One of the contribution comes from soft graviton connected to the external legs and the other comes from the  soft mode connected to the dressing. The soft graviton connected to  the external legs can be computed through leading single soft theorem in the undressed states and the other can be computed through contraction of the soft mode with the dressing operator $e^{R_{N}}$.
\\
External leg contribution can be computed through the leading single soft theorem in the undressed states \eqref{weinberg} as 
\begin{align}\label{sgol}
(\braout Q_{\mathrm{soft}}(\hat{p})\odress\sa\idress\ketin)_{\mathrm{external}}=\frac{\kappa}{2}S^{(0)}(\hat{p};\{k_{i}\})\braout\odress\sa\idress\ketin.
\end{align}
where $S^{(0)}(\hat{p};\{k_{i}\})$ is defined in\eqref{crazy-notation}.
\\\\
The contribution from the dressing can be evaluated as the contraction of the $Q_{\mathrm{soft}}(\hat{p})$ with the dressing operator i.e 
\begin{align}\nonumber\label{sgoc}
(\braout Q_{\mathrm{soft}}(\hat{p})\odress\sa\idress\ketin)_{\mathrm{dressing}}&=
\braout[Q_{\mathrm{soft}}(\hat{p}),\odress]\sa\idress\ketin+\\\nonumber&~~~~~~~~~~~~~~~~~~~~~~\braout\odress\sa[Q_{\mathrm{soft}}(\hat{p}),\idress]\ketin,\\
&=- \kappa \lim\limits_{E_{p}\rightarrow 0}E_{p} \Big(N_{\mathrm{out}}(p)-N_{\mathrm{in}}(p)\Big)\braout\odress\sa\idress\ketin.
\end{align}
In going from the first line to the second we used the commutation relations \eqref{cloudc}.
\\
Here $N_{\mathrm{out}}$ and $N_\mathrm{in}$ is  defined in \eqref{noutnin}
therefore the total contribution becomes the sum of \eqref{sgol} and \eqref{sgoc}. Hence,
\begin{align}\nonumber
\braout Q_{\mathrm{soft}}&(\hat{p})\odress\sa\idress\ketin=\\
&\kappa\Big(- \lim\limits_{E_{p}\rightarrow 0}E_{p} \Big(N_{\mathrm{out}}(p)-N_{\mathrm{in}}(p)\Big)+\frac{1}{2}S^{(0)}(\hat{p};\{k_{i}\})\Big)\braout\odress\sa\idress\ketin=0.
\end{align}
If we use the assumption \eqref{st3} we get,
\\
\begin{align}
\lim\limits_{E_{p}\rightarrow 0}E_{p} \Big(N_{\mathrm{out}}(p)-N_{\mathrm{out}}(p)\Big)=\frac{1}{2}S^{(0)}(\hat{p};\{k_{i}\}).
\end{align}
which is the same constraint as \eqref{sconstraint}.

To conclude, in perturbative gravity we can equivalently use the single orthogonality relation as well as single soft graviton theorem in the undressed states to constrain the dressing.
\section{Action of Hard Charge on Dressed States}\label{gra2}

In this section we give an argument for our assumption \eqref{st41}. We are interested in evaluating the following terms:
\begin{align}
-\braout[\odress,Q_{\mathrm{hard}}(\hat{p})]\sa\idress\ketin+\braout\odress\sa[Q_{\mathrm{hard}}(\hat{p}),\idress]\ketin.
\end{align}
%which can be expanded and written as
%\begin{align}
%\braout[\odress,\qfhard]\mathrm{S}&\idress\ketin+\braout\qfhard\odress\mathrm{S}\idress\ketin\\
%&-\braout\odress\mathrm{S}[\qfhard,\idress]\ketin-\braout\odress\mathrm{S}\idress\qfhard\ketin
%\end{align}
%We are mainly interested in the first and the third term which are
%\begin{align}
%\braout[\odress,\qfhard]\mathrm{S}\idress\ketin\\
%\braout\odress\mathrm{S}[\qfhard,\idress]\ketin
%\end{align}
%
%Consider
%\begin{align}
%\odress\qfhard\idress=\qfhard-[R_{f},\qfhard]+\frac{[R_{f},[R_{f},\qfhard]]}{2!}+...
%\end{align}
%Multiplying with $\odress$ on the right we get
%\begin{align}
%\odress\qfhard=\qfhard\odress-\Big([R_{f},\qfhard]-\frac{[R_{f},[R_{f},\qfhard]]}{2!}+..\Big)\odress
%\end{align}
The commutator in the first term of the above expression can be written as:
\begin{align}
[\odress,Q_{\mathrm{hard}}(\hat{p})]=-A\odress.
\end{align}
where
\begin{align}
A=\Big([R_{N}^{(\mathrm{out})},Q_{\mathrm{hard}}(\hat{p})]-\frac{[R_{N}^{(\mathrm{out})},[R_{N}^{(\mathrm{out})},Q_{\mathrm{hard}}(\hat{p})]]}{2!}+\ldots\Big).
\end{align}
Using the action of hard charge the first and second term in the above expression can be evaluated to
\begin{align}
[R_{N}^{(\mathrm{out})},Q_{\mathrm{hard}}(\hat{p})]&=-\int^{\Lambda} d[q]~S^{(0)}(\hat{p};q)~N^{\mu\nu}(q;k_{i})(a^{\dagger}_{\mu\nu}(q)+a_{\mu\nu}(q)),\\
[R_{N}^{(\mathrm{out})},[R_{N}^{(\mathrm{out})},Q_{\mathrm{hard}}(\hat{p})]]&=\int^{\Lambda} d[q]~S^{(0)}(\hat{p};q)N^{\mu\nu}(q;k_{i})N^{\rho\sigma}(q;k_{i})I_{\mu\nu\rho\sigma}.
\end{align}
where
\begin{align}
I_{\mu\nu\rho\sigma}\equiv \eta_{\mu\rho}\eta_{\nu\sigma}+\eta_{\mu\sigma}\eta_{\nu\rho}-\eta_{\mu\nu}\eta_{\rho\sigma}.
\end{align}
and $S^{(0)}(\hat{p};q)$ is already defined in \eqref{lfactor}.
It is not difficult to see that both of the integrals in the above expressions is of the order $\mathcal{O}(\Lambda)$ where $\Lambda$ is the upper cut off of the dressing operator. If the cutoff is to assumed significantly small so that only low energy gravitons constitute the dressing then these integrals are vanishing. Hence,
\begin{align}
[\odress,Q_{\mathrm{hard}}(\hat{p})]=\mathcal{O}(\Lambda).
\end{align}
\\
Similar analysis also holds for $[Q_{\mathrm{hard}}(\hat{p}),\idress]$.

%\begin{align}
%A=-\int \tilde{d^{3}k}~f_{k}E_{k}~N^{\mu\nu}(k)(a^{\dagger}_{\mu\nu}(k)+a_{\mu\nu}(k))-\frac{1}{2}\int \tilde{d^{3}k}~f_{k}E_{k}N^{\mu\nu}(k)N^{\rho\sigma}(k)I_{\mu\nu\rho\sigma}
%\end{align}

%Similarly multiplying with $\idress$ on the left we get
%\begin{align}
%\qfhard\idress=\idress\qfhard-\idress\Big([R_{f},\qfhard]-\frac{[R_{f},[R_{f},\qfhard]]}{2!}+..\Big)
%\end{align}
%Therefore
%\begin{align}
%[\qfhard,\idress]=-\idress A
%\end{align}
%This can be used to simplify the first and the third terms..
%\begin{align}
%[\odress,\qfhard]=-A\odress
%\end{align}
%As we found in the previous section the term $\frac{[R_{f},[R_{f},\qfhard]]}{2!}$ is just a c- number and doesnt involve any operators. Hence the extra term contributing from this can be written as 
%\begin{align}
%-\Big(\frac{1}{2}\int \tilde{d^{3}k}~&f_{k}E_{k}N^{\mu\nu}(k)N^{\rho\sigma}(k)I_{\mu\nu\rho\sigma}\Big)\braout\odress\sa\idress\ketin \nonumber \\
%&+	\Big(\frac{1}{2}\int \tilde{d^{3}k}~f_{k}E_{k}N^{\mu\nu}(k)N^{\rho\sigma}(k)I_{\mu\nu\rho\sigma}\Big)\braout\odress\sa\idress\ketin
%\end{align}
%Hence contribution from this term vanishes.\\
%The remaining terms are,
%\begin{align}
%\int \tilde{d^{3}k}~f_{k}E_{k}&~N^{\mu\nu}(k)\braout a_{\mu\nu}(k)\odress\sa\idress\ketin \cr
%&-\int \tilde{d^{3}k}~f_{k}E_{k}~N^{\mu\nu}(k)\braout\odress\sa\idress a^{\dagger}_{\mu\nu}(k)\ketin 
%\end{align}

\section{Action of Soft Charge on the Dressing Operator} \label{qcda1}
In this section we will derive the commutation relations of the soft gluon charge $Q^{a}_{\mathrm{soft}}(\hat{p})$ with the dressing operator. Let us first consider the dressing operator:
\begin{equation}\label{asd}
U_E(k)=\bar{P}_{E}\exp\Big(g\int^{E}{\mathrm{d}[q]} ~\mathcal{N}^{a \mu}(q;k)~A_{\mu}^{a}(q)\Big).
\end{equation} 
where, $A_{\mu}^{a}(q)=a_{\mu}^{a \dagger}(q)-a_{\mu}^{a}(q) $ and $ 
{\mathrm{d}[q]}=\frac{d^3 q}{(2 \pi)^3 (2 E_{q})}.
$
\\

%The creation and annihilation operators can be written in the polarisation basis,
%\begin{equation}
%a_{\mu}^{a \dagger}(q)=\sum_r \epsilon_{\mu}^r a^{a \dagger}(q)\quad \mathrm{and} \quad a_{\mu}^{a_1}(q)=\sum_r \epsilon_{\mu}^{r*} a^{a}(q)
%\end{equation}
As explained in section \ref{qcd32} due to the non-Abelian nature the operators in  \eqref{asd} are strictly ordered in energy.  The gluon operator with the lowest energy will act first. 
The $n$th order expansion of the dressing operator can be written as
\begin{equation}\label{eordering1}
U_{E}(k)_{{{\rvert}{g^n}}}=~g^n\int^{E}{\mathrm{d}[q_1]}\ldots\int^{E_{q_{n-1}}}{\mathrm{d}[q_n]}\mathcal{N}^{a_1 \mu_1}(q_{1};k)\ldots\mathcal{N}^{a_n \mu_n}(q_n;k) A_{\mu_{1}}^{a_1}(q_1)\ldots A_{\mu_{n}}^{a_n}(q_n).
\end{equation}

Let us consider the commutator of the soft operator $Q^\mathrm{soft}_{a}(\hat{p})$(defined as in \eqref{gs1}) on the $n$th order term in the dressing operator. i.e., we consider:
\begin{align}
[Q_\mathrm{soft}^{a}(\hat{p}),g^n\int^{E}{\mathrm{d}[q_1]}\ldots\int^{E_{q_{n-1}}}{\mathrm{d}[q_n]}\mathcal{N}^{a_1 \mu_1}(q_{1};k)\ldots\mathcal{N}^{a_n \mu_n}(q_n;k) A_{\mu_{1}}^{a_1}(q_1)\ldots A_{\mu_{n}}^{a_n}(q_n)].
\end{align}
Since the integral is ordered in energy with the lowest energy gluon operator on the right, the soft operator will act only on the right most operator. Hence
\begin{align}
[Q_\mathrm{soft}^{a}(\hat{p}),U_{E}(k)_{{{\rvert}{g^n}}}]=gU_{E}(k)_{{{\rvert}{g^{n-1}}}}\lim\limits_{E_p \to{ 0}}~E_p\mathcal{N}^{a \mu}(p;k)\epsilon^{+}_{\mu}(\hat{p}).
\end{align}
It is now easy to see that
\begin{align}\label{mswi1}
[Q_\mathrm{soft}^{a}(\hat{p}),U_{E}(k)]=g U_{E}(k)\lim\limits_{E_p \to{ 0}}E_p ~\mathcal{N}^{a \mu}(p;k)\epsilon_{\mu}^+(\hat{p}).
\end{align}
Similarly one can also see that
\begin{align}\label{mswi2}
[Q_\mathrm{soft}^{a}(\hat{p}),U_{E}^{\dagger}(k)]=-g\lim\limits_{E_p \to{ 0}}E_p ~\mathcal{N}^{a \mu}(p;k)\epsilon_{\mu}^+(p)U_{E}^{\dagger}(k).
\end{align}
\\
If we now consider the modified dressing operator defined as
\begin{align}\nonumber
U_{E}(k)=\bar{P}_{E}\exp\Big(g \int^{E}&{\mathrm{d}[q_1]} \mathcal{N}_{(1)}^{a_1 \mu_1}(q_1;k) A_{\mu_{1}}^{a_1}(q_1)+\\&g^2 \int^{E}{\mathrm{d}[q_1]}\int^{E_{q_1}}{\mathrm{d}[q_2]}\mathcal{N}_{(2)}^{a_2 \mu_2,a_1 \mu_1}(q_{1},q_{2};k)A_{\mu_{1}}^{a_1}(q_1)A_{\mu_{2}}^{a_2}(q_2)+\ldots\Big).
\end{align}
%Let us denote the terms inside the exponential by $R_{N}$
%Therefore
%\begin{align}
%U_{E}(k)=\bar{P}_{E}\exp(R_{N})
%\end{align}
%where 
%\begin{align}\nonumber
%R_{N}\equiv\int^{E}{\mathrm{d}[q_1]} \mathcal{N}_{(1)}^{a_1 \mu_1}&(q_1;k) A_{\mu_{1}}^{a_1}(q_1)\\
%&+g^2 \int^{E}{\mathrm{d}[q_1]}\int^{E_{q_1}}{\mathrm{d}[q_2]}\mathcal{N}_{(2)}^{a_2 \mu_2,a_1 \mu_1}(q_{1},q_{2};k)A_{\mu_{1}}^{a_1}(q_1)A_{\mu_{2}}^{a_2}(q_2)+...
%\end{align}
\\
One could use the same analysis for the commutation relation of the soft operator with the modified dressing operator and verify that
\begin{align}\label{mswi3}
[Q_\mathrm{soft}^{a}(\hat{p}),U_{E}(k)]=g ~U_{E}(k)\lim\limits_{E_p \to{ 0}}E_p ~\mathcal{N}^{a \mu}(p;k)\epsilon_{\mu}^{+}(\hat{p}) +\mathcal{O}(g^{2})+\ldots
\end{align}
Using this result we can also find the nested commutator.
\begin{align}\label{mswi4} \nonumber
[Q_\mathrm{soft}^{a}(\hat{p_{1}}),[Q_\mathrm{soft}^{b}&(\hat{p_2}),U_{E}(k)]]=g^{2}U_{E}(k)\lim\limits_{E_{p_{2}} \to{ 0}}E_{p_{2}} \lim\limits_{E_{p_{1}} \to{ 0}}E_{p_{1}} \mathcal{N}^{a \mu}(p_{1};k)\mathcal{N}^{b \nu}(p_{2};k)\epsilon^+_{\mu}(\hat{p}_1)\epsilon^{+}_{\nu}(\hat{p}_2)\\&+g^{2}U_{E}(k)\lim\limits_{E_{p_{2}} \to{ 0}}E_{p_{2}} \lim\limits_{E_{p_{1}} \to{ 0}}E_{p_{1}}\mathcal{N}_{(2)}^{a \mu, b \nu}(p_1,p_2;k)\epsilon^+_{\mu}(\hat{p}_1)\epsilon^+_{\nu}(\hat{p}_2)+ \mathcal{O}(g^{3}).
\end{align}

%\section{Action of gluon hard charge on the dressing operator}\label{qcda2}
%
%In this section we will argue that  using color conservation, 
%
%\begin{equation}
%\langle\textrm{out}\vert [Q, U_{E}^{h}]S\ -\ S[Q, U_{E}^{h}]\vert \Omega\rangle\ =\ 0 \textrm{at}\ O(g)
%\end{equation}
%
%For simplicity, we take all the quarks to be outgoing and take the incoming state to be the (undressed) vacuum $\vert\Omega\rangle$. 
%
%It is a simple exercise to show that at the leading order in $g$
%
%\begin{equation}
%\langle\textrm{out}\vert [Q, U_{E}^{h}]S\ -\ S[Q, U_{E}^{h}]\vert \Omega\rangle\ =\\
%\sum_{i,j}\int d\mu(k) \frac{p_{i}\cdot p_{j}\ -\ (p_{i}\cdot\hat{k})(p_{j}\cdot\hat{k})}{(p_{i}\cdot k)(p_{j}\cdot k)}\hat{T}^{a}_{i}T^{b}_{j}f^{abc}\textrm{out}\vert S \vert\Omega\rangle\\
%=\ \sum_{i\neq j}\dots 
%\end{equation}

\section{Nested Ward Identity with Dressed States}\label{qcda3}
%In the previous section we have obtained the generalised coherent state. And we have seen that with such a dressed state, soft modes always decouples from the S matrix. We now ask if this new dressing is indicative of a Hierarchy of conservation laws?.  We have already checked the consistency of single ward identity with generalised dressed state. But when we consider multiple charges there exists difficulties at the level of ward identity itself. We will discuss this in the context of double ward identity.  
%\begin{align}\nonumber
%U_{E}(k)=\bar{P}_{\omega}\exp\Big(g \int^{E}{\mathrm{d}[q_1]} & \mathcal{N}_{(1)}^{a_1 \mu_1}(q_1;k) A_{\mu_{1}}^{a_1}(q_1)+\\&g^2 \int^{E}{\mathrm{d}[q_1]}\int^{E_{q_1}}{\mathrm{d}[q_2]}\mathcal{N}_{(2)}^{a_2 \mu_2,a_1 \mu_1}(q_{1},q_{2};k)A_{\mu_{1}}^{a_1}(q_1)A_{\mu_{2}}^{a_2}(q_2)\Big)
%\end{align}
%By writing the LGT charge as a sum of part $ Q_{\mathrm{soft}}^a(\hat{p})$ and hard $Q_{\mathrm{hard}}^a(\hat{p}) $ we will get,
%\begin{align}\nonumber
%{}_{\mathrm{d}}\bra{\mathrm{out}}\Big([Q_{\mathrm{soft}}^a(\hat{p_{1}})],[Q_{\mathrm{soft}}^b(\hat{p_{2}}),\mathrm{S}]]+[Q_{\mathrm{soft}}^a(\hat{p_{1}}),[Q_{\mathrm{hard}}^b(\hat{p_{2}}),&\mathrm{S}]]+[Q_{\mathrm{hard}}^a(\hat{p_{1}}),[Q_{\mathrm{soft}}^b(\hat{p_{2}}),\mathrm{S}]]\\ &+[Q_{\mathrm{hard}}^a(\hat{p_{1}}),[Q_{\mathrm{hard}}^b(\hat{p_{2}}),\mathrm{S}]]\Big)\ket{\mathrm{in}}_{\mathrm{d}}=0.\label{QCD_double_WI}
%\end{align}
In this section we evaluate \eqref{dwi2} using the action of the charges. We start with the l.h.s of \eqref{dwi2}
\begin{align}\label{main}
\braout\qout[Q^{a}_{\mathrm{soft}}(\hat{p}_{1})+Q^{a}_{\mathrm{hard}}(\hat{p}_{1}),[Q^{b}_{\mathrm{soft}}(\hat{p}_{2})+Q^{b}_{\mathrm{hard}}(\hat{p}_{2}),S]]\qin\ketin.
\end{align}
which can be written as
\begin{align}\nonumber\label{dwi21}
&\braout\qout[Q^{a}_{\mathrm{soft}}(\hat{p}_{1}),[Q^{b}_{\mathrm{soft}}(\hat{p}_{2})+Q^{b}_{\mathrm{hard}}(\hat{p}_{2}),S]]\qin\ketin\\&~~~~~~~~~~~~~~~~~~~~~~~~~~~~~~
 +\braout\qout[Q^{a}_{\mathrm{hard}}(\hat{p}_{1}),[Q^{b}_{\mathrm{soft}}(\hat{p}_{2})+Q^{b}_{\mathrm{hard}}(\hat{p}_{2}),S]]\qin\ketin.
\end{align}
\\
Before proceeding with the computation, we note that the action of $Q_{\textrm{hard}}^{a}(\hat{p})$ on the dressing operator will produce terms of $O(g^{2})$. This together with the action of $Q_{\textrm{soft}}^{a}(\hat{p})$ on the external states or the dressing operator will contribute at $O(g^{3})$. Therefore the terms involving the action of $Q_{\textrm{hard}}^{a}(\hat{p})$ on $\qout$ or $\qin$ in \ref{dwi21} will not contribute at $O(g^{2})$.  \\
Let us consider the second term in the above equation first. Since the hard charge will only act on the external states this becomes
\begin{align}\nonumber
\braout\qout[Q^{a}_{\mathrm{hard}}(\hat{p}_{1}),[Q^{b}_{\mathrm{soft}}(\hat{p}_{2})+Q^{b}_{\mathrm{hard}}(\hat{p}_{2}),S]]\qin\ketin&=gS^{(0)a}(\hat{p}_{1};\{k_{i}\})\braout\qout[Q^{b}(\hat{p}_{2}),\sa]\qin\ketin,\\
&=0.
\end{align}
In going from the first line to the second we used the Ward identity involving one asymptotic charge. Therefore in \eqref{dwi21} we are left with evaluating
\begin{align}\nonumber\label{dwi4}
\braout&\qout[Q^{a}_{\mathrm{soft}}(\hat{p}_{1}),[Q^{b}_{\mathrm{soft}}(\hat{p}_{2})+Q^{b}_{\mathrm{hard}}(\hat{p}_{2}),S]]\qin\ketin\\&= \braout\qout[Q^{a}_{\mathrm{soft}}(\hat{p}_{1}),[Q^{b}_{\mathrm{hard}}(\hat{p}_{2}),\sa]]\qin\ketin +\braout\qout[Q^{a}_{\mathrm{soft}}(\hat{p}_{1}),[Q^{b}_{\mathrm{soft}}(\hat{p}_{2}),\sa]]\qin\ketin.
\end{align}
\\
The first term in the above expression can be expanded as
\begin{align}\nonumber\label{dwi3}
&\braout\qhb[\qout,\qsa]\sa\qin\ketin-\braout[\qout,\qsa]\sa\qin\qhb\ketin-\\\nonumber
&~~\braout\qhb\qout\sa[\qsa,\qin]\ketin+\braout\qout\sa[\qsa,\qin]\qhb\ketin+\\
&~~~~~~~~~\braout\qout[\qsa,\qhb]\sa\qin\ketin+\braout\qout\sa[\qhb,\qsa]\qin\ketin.
\end{align}
which can be evaluated to 
\begin{align}\nonumber
\braout&\qout[Q^{a}_{\mathrm{soft}}(\hat{p}_{1}),[Q^{b}_{\mathrm{hard}}(\hat{p}_{2}),\sa]]\qin\ketin\\\nonumber
&=-\Big(g^{2} \lim\limits_{E_{p_1} \to 0} E_{p_1}\mathcal{N}_{(1)}^{a \mu}(p_1;\{k_j\})\epsilon_{\mu}^{+}(p_1)S^{(0)b}(\hat{p_2};\{k_{i}\})\\\nonumber&~~~~~~~~~~~~+\lim\limits_{E_{p_2} \to 0} E_{p_2}( S^{(0)b}(\hat{p}_{2};p_{1}))_{ac}~\mathcal{N}_{(1)}^{c \mu}(p_1;\{k_i\})\epsilon_{\mu}^{+}(p_2)\Big)\bra{\mathrm{out}}U_{E}^{\dagger\mathrm{out}}SU_{E}^{\mathrm{in}} \ket{\mathrm{in}}+\mathcal{O}(g^{3}),\\\nonumber
&=-g^{2}\Big( S^{(0)a}(\hat{p_1};\{k_{j}\})S^{(0)b}(\hat{p_2};\{k_{i}\})\\&~~~~~~~~~~~~~~~~~~~~~~~~~~~~~~~+( S^{(0)b}(\hat{p}_{2};p_{1}))_{ac}~S^{(0)c}(\hat{p_1};\{k_{i}\})\Big)\bra{\mathrm{out}}U_{E}^{\dagger\mathrm{out}}SU_{E}^{\mathrm{in}} \ket{\mathrm{in}}+\mathcal{O}(g^{3}).
\label{softhard1}
\end{align}
\\
For evaluating the last two terms in the expression \eqref{dwi3} we have used the commutator
\begin{align}\label{hsc}
[\qsa,\qhb]=-g(S^{(0)b}(\hat{p}_{2};p_{1}))_{ac}~Q_{\mathrm{soft}}^{c}(\hat{{p_{1}}}).
\end{align}
\ref{hsc} can be easily derived using the definition of $\qsa$ and $\qhb$ in \ref{gs1} and \ref{gh1}\footnote{It is important to note that this is valid only at tree-level. The hard charge recieves loop corrections starting at $\mathcal{O}(g^{2})$ and therefore therefore \ref{hsc} recieves loop-corrections. But these will not affect our analysis since such contributions to \ref{main} will be at $\mathcal{O}(g^{3})$ and higher.}. 
\begin{align}\nonumber
[\qsa,\qhb]&=\Big[\lim\limits_{E_{p} \to{ 0}}E_{p_{1}}~ a_{+}^{a}(E_{p_{1}}~\hat{{p_{1}}}),~\qhb\Big]\\\nonumber
&=\lim\limits_{E_{p} \to{ 0}}E_{p_{1}}\Big[a_{+}^{a}(E_{p_{1}}~\hat{{p_{1}}}),~\qhb\Big]\\\nonumber
&=-\lim\limits_{E_{p_{1}} \to{ 0}}E_{p_{1}}~g~\Big(S^{(0)b}(\hat{{p_{2}}},p_{1})\Big)_{ac}~a_{+}^{c}(E_{p_{1}}~\hat{{p_{1}}})\\
&=-g~\Big(S^{(0)b}(\hat{{p_{2}}},p_{1})\Big)_{ac}~Q_{\mathrm{soft}}^{c}(\hat{{p_{1}}})
\label{softhardc1}
\end{align}\\
%\begin{align}\nonumber
%[\qsa,\qhb]\ket{(k,c)}&=\qsa\qhb\ket{(k,c)}-\qhb\qsa\ket{(k,c)}\\\nonumber
%&=-g \qsa S^{(0)b}(\hat{{p_{2}}};k)_{cd}\ket{(k,d)}-\qhb\lim\limits_{E_{p} \to{ 0}}E_{p_{1}}~ a_{+}^{a}(E_{p_{1}}~\hat{{p_{1}}})\ket{(k,c)}\\\nonumber
%&=-g \qsa S^{(0)b}(\hat{{p_{2}}};k)_{cd}\ket{(k,d)}-gS^{(0)b}(\hat{{p_{2}}};p_{1})_{ad}\lim\limits_{E_{p} \to{ 0}}E_{p_{1}}~ a_{+}^{d}(E_{p_{1}}~\hat{{p_{1}}})\ket{(k,c)}\\\nonumber
%&~~~~~~~~~~~~~~~~~~~~~~~~~~~~~~~~~~~~~~~~~~+g \qsa S^{(0)b}(\hat{{p_{2}}};k)_{cd}\ket{(k,d)}\\\nonumber
%&=-gS^{(0)b}(\hat{{p_{2}}};p_{1})_{ad}\lim\limits_{E_{p} \to{ 0}}E_{p_{1}}~ a_{+}^{d}(E_{p_{1}}~\hat{{p_{1}}})\ket{(k,c)}\\
%&=-gS^{(0)b}(\hat{{p_{2}}};p_{1})_{ad}~~Q_{\mathrm{soft}}^{d}(\hat{{p_{1}}})\ket{(k,c)}
%\label{softhardc}
%\end{align}\\
Similarly using orthogonality condition involving one soft gluon mode the second term in \eqref{dwi4} reduces to
\begin{align}\nonumber
\braout[[\qout,\qsa],\qsb]\sa\qin\ketin+\braout\qout\sa[\qsb,[\qsa,\qin]]\ketin-\\\nonumber
\braout[\qout,\qsa]\sa[\qsb,\qin]\ketin-\braout[\qout,\qsb]\sa[\qsa,\qin]\ketin\\
~~~~~~+\braout\qsa\qsb\qout\sa\qin\ketin.
\end{align}
\\
The commutators in the given expression has already been derived in appendix \ref{qcda1} (\eqref{mswi3} and \eqref{mswi4}). Hence the above expression can be  evaluated to be
\begin{align}\nonumber\label{s11}
&g^2\lim\limits_{E_{p_2} \to 0} E_{p_2}\lim\limits_{E_{p_1} \to 0} E_{p_1}\Big(\mathcal{N}_{(1)}^{a \mu}(p_1;\{k_j\})\mathcal{N}_{(1)}^{b \nu}(p_2;\{k_i\})+\mathcal{N}_{(2)}^{a \mu,b \nu }(p_1,p_2;\{k_i\})\Big)\epsilon_{\mu}^{+}(p_1)\epsilon_{\nu}^{+}(p_2)\\\nonumber &~~~~~~~~~~~~~~~~~\times \bra{\mathrm{out}}U_{E}^{\dagger\mathrm{out}}SU_{E}^{\mathrm{in}} \ket{\mathrm{in}}+\braout\qsa\qsb\qout\sa\qin\ketin\\\nonumber
&= g^2 \Big(S^{(0)a}(\hat{p_1};\{k_{j}\})S^{(0)b}(\hat{p_2};\{k_{i}\}) + \lim\limits_{E_{p_2} \to 0} E_{p_2}\lim\limits_{E_{p_1} \to 0} E_{p_1}\mathcal{N}_{(2)}^{a \mu,b \nu }(p_1,p_2;\{k_i\})\epsilon_{\mu}^{+}(\hat{p}_1)\epsilon_{\nu}^{+}(\hat{p}_2)\Big) \\
&~~~~~~~~~~~~~\times \bra{\mathrm{out}}U_{E}^{\dagger\mathrm{out}}SU_{E}^{\mathrm{in}} \ket{\mathrm{in}}+\braout\qsa\qsb\qout\sa\qin\ketin+\mathcal{O}(g^{3}).
\end{align}

where
\begin{align}
\mathcal{N}_{(2)}^{a \mu,b \nu }(p_1,p_2;\{k_i\})=\sum_{i=\mathrm{out}}\mathcal{N}_{(2)}^{a \mu,b \nu }(p_1,p_2;k_i)-\sum_{i=\mathrm{in}}\mathcal{N}_{(2)}^{a \mu,b \nu }(p_1,p_2;k_i).
\end{align}

Therefore summing the contributions \eqref{softhard1} and \eqref{s11} we finally get \eqref{main} as
\begin{align}\nonumber
g^2 ~\Big( \lim\limits_{E_{p_2} \to 0} E_{p_2}\lim\limits_{E_{p_1} \to 0} E_{p_1}\mathcal{N}_{(2)}^{a \mu,b \nu }(p_1,p_2;\{k_i\})\epsilon_{\mu}^{+}(\hat{p}_1)\epsilon_{\nu}^{+}(\hat{p}_2)- S^{(0)b}(\hat{p}_{2};p_{1}))_{ac}~S^{(0)c}(\hat{p_1};\{k_{i}\})\Big)\\
\times \bra{\mathrm{out}}U_{E}^{\dagger\mathrm{out}}SU_{E}^{\mathrm{in}} \ket{\mathrm{in}}+\braout\qsa\qsb\qout\sa\qin\ketin+\mathcal{O}(g^{3}).
\end{align}

\end{document}